\documentclass[%
 reprint,
 amsmath,amssymb,
 aps,
]{revtex4-2}

\usepackage{graphicx}
\usepackage{bm}
\usepackage{color}
\usepackage{booktabs}

\begin{document}

\preprint{APS/123-QED}

\title{Sourced stationary Dirac bound states around Kerr black holes}

\author{Sen Guo$^{1}$, Peng-Yu Chen$^{1,2}$, Yi-Han Huang$^{1}$, Xin Li$^{1}$, Yu Liang$^{3}$, Kai Lin$^{4}$, Lin Wen$^{1*}$, Li-Fang Li$^{5*}$}
\address{
$^1$College of Physics and Optoelectronic Engineering, Chongqing Normal University, Chongqing 401331, People's Republic of China\\
$^2$School of Physics and Astronomy, China West Normal University, Nanchong 637000, People's Republic of China\\
$^3$School of Big Data and Artificial Intelligence, Fuyang University of Technology, Fuyang 236000, People's Republic of China\\
$^4$Universidade Federal de Campina Grande, Campina Grande, PB, Brasil, Instituto de F\'isica, Universidade de S\~ao Paulo 05508-900, S\~ao Paulo, Brasil\\
$^5$Center for Gravitational Wave Experiment, National Microgravity Laboratory, Institute of Mechanics, Chinese Academy of Sciences, Beijing 100190, People's Republic of China}
\email{$^*$ lilifang@imech.ac.cn; $^*$ wlqx@cqnu.edu.cn (Corresponding author)}



\date{\today}

\begin{abstract}
Ultralight bosonic fields can form macroscopic clouds around rotating black holes, whereas the existence of analogous stationary fermionic configurations is constrained by their intrinsic spin. Here we study stationary bound states of massive Dirac fields supported by an external source on Kerr backgrounds, in the test-field approximation, with a kinematic extension to Kerr--Newman, they are not fermionic condensates in the many-body sense. By mapping the Kerr--Dirac system to a globally integrated sourced radial problem, we isolate the boundary constraints dictated by horizon causality. The angular sector shows that for $|m|=1/2$ the regular boundary branches prevent the field density from vanishing on the rotation axis, whereas modes with $|m|\ge3/2$ vanish there, the reconstructed Kerr density is only mildly oblate. Precisely at the kinematic synchronization locus, the Frobenius matrix of the Dirac operator is non-defective and devoid of logarithmic divergences. Regularity on the future horizon nevertheless selects the ingoing solution, and an exact radial flux identity shows that this solution carries a positive flux, which excludes source-free normalizable states at all frequencies. The exclusion of source-free synchronized states demonstrates that a black hole's capacity to support macroscopic stationary fields is governed not merely by superradiant kinematics, but by the interplay between horizon regularity and the positivity of the Dirac flux.

\vspace{1em}
\end{abstract}

\maketitle


\section{Introduction}
\label{sec:intro}
\par
In general relativity, the event horizon is a one-way causal boundary. At the quantum level, however, this causal structure distorts the Minkowski vacuum, altering the local mode spectrum~\cite{Hawking:1975vcx}. This vacuum polarization manifests as Hawking radiation, linking the black hole's thermodynamic temperature to its surface gravity, and its entropy to the horizon area~\cite{Bekenstein:1973ur,Bardeen:1973gs}. Such geometrical-thermodynamic correspondence suggests the near-horizon geometry can catalyze other non-trivial quantum phenomena. For instance, generalized uncertainty principles in this strong-field regime enforce a minimal Planck-scale spatial resolution~\cite{Maggiore:1993kv,Capozziello:1999wx}.

\par
The formulation of the Dirac field in rotating black holes relies on Chandrasekhar's exact separability in the Kerr metric and its extension to the Kerr--Newman metric by Page~\cite{Chandrasekhar:1976ap,Page:1976jj,Chandrasekhar:1984siy}. Beyond the test-field regime, the non-linear Einstein--Dirac system admits self-gravitating spinor bound states~\cite{Finster:1998ws,Herdeiro:2017fhv,Herdeiro:2019mbz}, and the condensation of a classical spinor field at the horizon of a self-gravitating Einstein--Dirac configuration has recently been discussed~\cite{Dzhunushaliev:2025dma}. However, the stationary spectrum of the Dirac field is tightly constrained by rigorous no-go theorems. Finster et al. proved that normalizable, time-periodic Dirac solutions do not exist for non-extreme Reissner--Nordstr\"om black holes~\cite{Finster:1998ak} and, more generally, for non-extreme Kerr--Newman and separable stationary axisymmetric black holes~\cite{Finster:1999ry}; Dirac waves in the Kerr--Newman exterior decay in time~\cite{Finster:2000jz}, and static, spherically symmetric Einstein--Dirac--Maxwell black holes with spinor hair do not exist~\cite{Finster:1998ju}. Massive Dirac fields around Schwarzschild and Kerr black holes therefore have only decaying quasi-bound states~\cite{Lasenby:2002mc,Dolan:2015eua}. This restriction is circumvented only in the extremal limit at exact corotation, and then only for discrete parameter values~\cite{Schmid:2002zf}, although the existence of these states has been questioned. Such fermionic behavior contrasts sharply with the bosonic sector, where rotational superradiance sustains source-free stationary scalar clouds, driving the formation of hairy black holes~\cite{Hod:2012px,Herdeiro:2014goa,Guo:2026pdi}. All of these fermionic results, however, concern the homogeneous Dirac equation.

\par
It has recently been proposed that near-horizon fermionic anticommutation relations might deviate from their flat-spacetime limits~\cite{Dzhunushaliev:2026rrt}. This was phenomenologically modeled by injecting a macroscopic source into the equal-time bilocal Dirac equation. By tuning this source to cancel singular static-frame coefficients while enforcing spatial decay, stationary configurations with finite static-frame amplitudes were found on a Schwarzschild background. Yet, this framework has clear limitations: it prescribes the source manually rather than deriving it from a deformed field algebra, and it bypasses the surface gravity---the core thermodynamic parameter dictating near-horizon quantum states~\cite{McMaken:2024tpc,Dubey:2025hwk}. Furthermore, the singular coefficients requiring regularization are largely artifacts of the static orthonormal frame, whose proper acceleration diverges at the horizon~\cite{Wald:1984rg}; freely falling observers, by contrast, measure a regular local effective temperature~\cite{McMaken:2024tpc}. Finally, in standard quantum field theory in curved spacetime the two-point (Wightman) functions of the Dirac field are bisolutions of the homogeneous Dirac equation, and inhomogeneous terms appear only for propagators, as $\delta$-function sources at coincident points~\cite{Birrell:1982ix,Fewster:2025vxz}; a source extended over the exterior therefore lies outside the standard framework.

\par
Extending this analysis to the Kerr geometry disentangles frame-dependent artifacts from the invariant near-horizon dynamics, characterized by a dimensionless geometric index~\cite{Chandrasekhar:1984siy}. For non-extremal black holes, this index is parameterized by the surface gravity $\kappa_+$ (and thus the Hawking temperature $T_H = \kappa_+ / (2\pi)$) and the kinematic detuning $\omega - m\Omega_H$~\cite{Page:1976jj}. In Boyer--Lindquist coordinates its imaginary part produces an unbounded radial phase winding near the horizon. For the ingoing local solution this winding is a coordinate artifact that disappears in ingoing Kerr coordinates, whereas the outgoing solution is singular on the future horizon. Regularity therefore selects the ingoing solution at every frequency, and the obstruction to source-free bound states is global rather than local~\cite{Chandrasekhar:1976ap,Teukolsky:1974yv}. In the non-extremal regime, this winding vanishes only at exact corotation ($\omega = m\Omega_H$), mirroring the synchronization condition for bosonic clouds~\cite{Hod:2012px}; for each half-integer $m$ this fixes a single frequency. Nevertheless, satisfying this local kinematic constraint remains insufficient to guarantee a globally regular, source-free bound state.

\par
Physically, the exclusion of source-free, time-periodic bound states on non-extremal backgrounds~\cite{Finster:1998ak,Finster:1999ry} is tied to the causal character of the Dirac current, which makes the flux density across the future horizon non-negative for every configuration regular there; the same property underlies the absence of Dirac superradiance~\cite{Unruh:1974bw,Martellini:1977qf,Chandrasekhar:1984siy}. For a stationary bound state the integrated horizon flux must vanish. A non-negative flux density can vanish, however, not only when the field vanishes on the horizon but also when the current is null and tangent to the generators; in Sec.~\ref{sec:3.2} we use an exact radial flux identity to show that, for regular homogeneous solutions, the field must then vanish identically. The two sourced families of Ref.~\cite{Dzhunushaliev:2026rrt} instead prescribe non-vanishing (Case A) or vanishing (Case B) limits of the static-frame radial amplitudes. These exterior limits do not imply a smooth field on the future horizon: Case A is singular in a horizon-regular frame, and Case B is continuous but not differentiable there (Sec.~\ref{sec:3.2}).

\par
The corotation regime highlights the structural dichotomy between bosonic and fermionic stationary states. In the non-extremal Dirac problem, while the two Frobenius indices coalesce at corotation, the leading coefficient matrix remains non-defective~\cite{CoddingtonLevinson:1955,Wasow:1965}. This yields two linearly independent local horizon solutions without a logarithmic branch. For scalar fields the coalescence produces a logarithmic second solution, which horizon regularity excises~\cite{CoddingtonLevinson:1955,Hod:2012px}. Since regularity also excludes the outgoing Dirac solution, the two cases differ in the flux carried by the selected solution: for scalars it is proportional to $\omega-m\Omega_H$ and vanishes at synchronization, which permits clouds for discrete parameters~\cite{Hod:2012px,Herdeiro:2014goa}, whereas for Dirac fields it is positive, and synchronization offers no loophole~\cite{Finster:1999ry}. Although this bound-state obstruction shares physical roots with the absence of fermionic superradiance, their domains differ: the former governs localized stationary clouds, while the latter dictates asymptotic scattering amplitudes~\cite{Unruh:1974bw,Martellini:1977qf,Brito:2015rjv}.

\par
We work in the test-field approximation with an external source: the Dirac field is a c-number spinor on a fixed Kerr (or Kerr--Newman) background that obeys a linear Dirac equation with a prescribed source, and self-interactions, self-gravity, and backreaction are neglected. The calculation therefore does not describe a fermionic condensate in the many-body sense, and we call its solutions sourced stationary Dirac bound states. In Ref.~\cite{Dzhunushaliev:2026rrt} the solution is interpreted as an equal-time two-point function $\langle Q|\hat\psi^\dagger(t,\vec r_H)\hat\psi(t,\vec r)|Q\rangle$ in a state $|Q\rangle$ and, because it is stationary, as a horizon fermion condensate. Our calculation uses neither the anticommutation relations nor a quantum state and is mathematically the same as for an externally maintained classical spinor field. Our results rest on classical properties of the Dirac equation (its first-order form, the half-integer azimuthal number, and the positivity of the Dirac current~\cite{Dolan:2015eua}); they are compatible with spin statistics but do not test it.

\par
In this work, we generalize the sourced Dirac framework of Ref.~\cite{Dzhunushaliev:2026rrt} to the Kerr spacetime via the Chandrasekhar--Page separation scheme~\cite{Chandrasekhar:1984siy,Page:1976jj}. We study how rotation modifies the near-horizon regularity conditions for sourced Dirac bound states, quantify the requisite source strength, and trace the evolution of the solution branches with black hole spin. Section~\ref{sec:2} formulates the problem, using the Kinnersley null tetrad to decouple the Dirac equation, followed by a Frobenius analysis to extract the phase winding and extremal asymptotics; Sec.~\ref{sec:2.3} discusses the origin of the source and which of its ingredients are postulated. In Section~\ref{sec:3}, we isolate the removable coordinate artifact of the Schwarzschild horizon, benchmark against known eigenfrequencies, specify the frames used to assess regularity, and derive an exact flux identity that excludes source-free bound states. Section~\ref{sec:4} presents the angular eigenvalue spectrum, a full reconstruction of the field density, and self-consistent numerical integrations of the radial system, continued through turning points. These are validated in the vanishing-spin limit before mapping the frequency--spin parameter space, probing the corotation regime, testing the sensitivity to the source profile (Sec.~\ref{sec:4.3}), and extending to the Kerr--Newman geometry. Section~\ref{sec:5} summarizes our conclusions, and Appendix~\ref{app:repro} gives reproducibility data.

\section{Sourced Dirac equation in the Kerr spacetime}
\label{sec:2}
\par
\subsection{Geometric framework and radial separation}
\label{sec:2.1}
\par
The Kerr metric~\cite{Kerr:1963ud} for a rotating black hole of mass $M$ and angular momentum $J=aM$ is given in Boyer--Lindquist coordinates by
\begin{eqnarray}
ds^{2}=&-\Big(1-\frac{2Mr}{\Sigma}\Big)dt^{2}
-\frac{4Mar\sin^{2}\theta}{\Sigma}\,dt\,d\varphi
+\frac{\Sigma}{\Delta}\,dr^{2}+\Sigma\,d\theta^{2} \nonumber\\
&+\Big(r^{2}+a^{2}+\frac{2Ma^{2}r\sin^{2}\theta}{\Sigma}\Big)\sin^{2}\theta\,d\varphi^{2},
\label{eq-2.1}
\end{eqnarray}
where
\begin{equation}
\Delta(r)=r^{2}-2Mr+a^{2},\qquad
\Sigma(r,\theta)=r^{2}+a^{2}\cos^{2}\theta ,
\label{eq-2.2}
\end{equation}
and $\sqrt{-g}=\Sigma\sin\theta$. The essential departure from spherical symmetry is the off-diagonal term $g_{t\varphi}=-2Mar\sin^{2}\theta/\Sigma$, which induces frame-dragging~\cite{Misner:1973prb}. The roots $\Delta=0$ give the inner and outer horizons,
\begin{equation}
r_{\pm}=M\pm\sqrt{M^{2}-a^{2}},\qquad \Delta=(r-r_{+})(r-r_{-}),
\label{eq-2.3}
\end{equation}
satisfying $r_{\pm}^{2}+a^{2}=2Mr_{\pm}$. The ergosphere, where the time-translation Killing vector $\partial_t$ becomes null, is located at $r_{\rm ergo}(\theta)=M+\sqrt{M^{2}-a^{2}\cos^{2}\theta}$~\cite{Wald:1984rg}. Within this region, $\partial_t$ is spacelike, providing the kinematic prerequisite for bosonic superradiance~\cite{Zeldovich:1971ffh}. The outer horizon is governed by its angular velocity $\Omega_{H}$ and surface gravity $\kappa_{+}$~\cite{Bardeen:1973gs}:
\begin{eqnarray}
\Omega_{H}&=-\frac{g_{t\varphi}}{g_{\varphi\varphi}}\bigg|_{r_{+}}
=\frac{a}{r_{+}^{2}+a^{2}}=\frac{a}{2Mr_{+}} ,
\label{eq-2.4}\\
\kappa_{+}&=\frac{r_{+}-r_{-}}{2(r_{+}^{2}+a^{2})}
=\frac{\sqrt{M^{2}-a^{2}}}{2M\big(M+\sqrt{M^{2}-a^{2}}\,\big)} ,
\label{eq-2.5}
\end{eqnarray}
which defines the Hawking temperature $T_{H}=\kappa_{+}/2\pi$~\cite{Hawking:1975vcx}.

\par
\begin{figure*}[htbp]
\centering
\includegraphics[width=12.5cm]{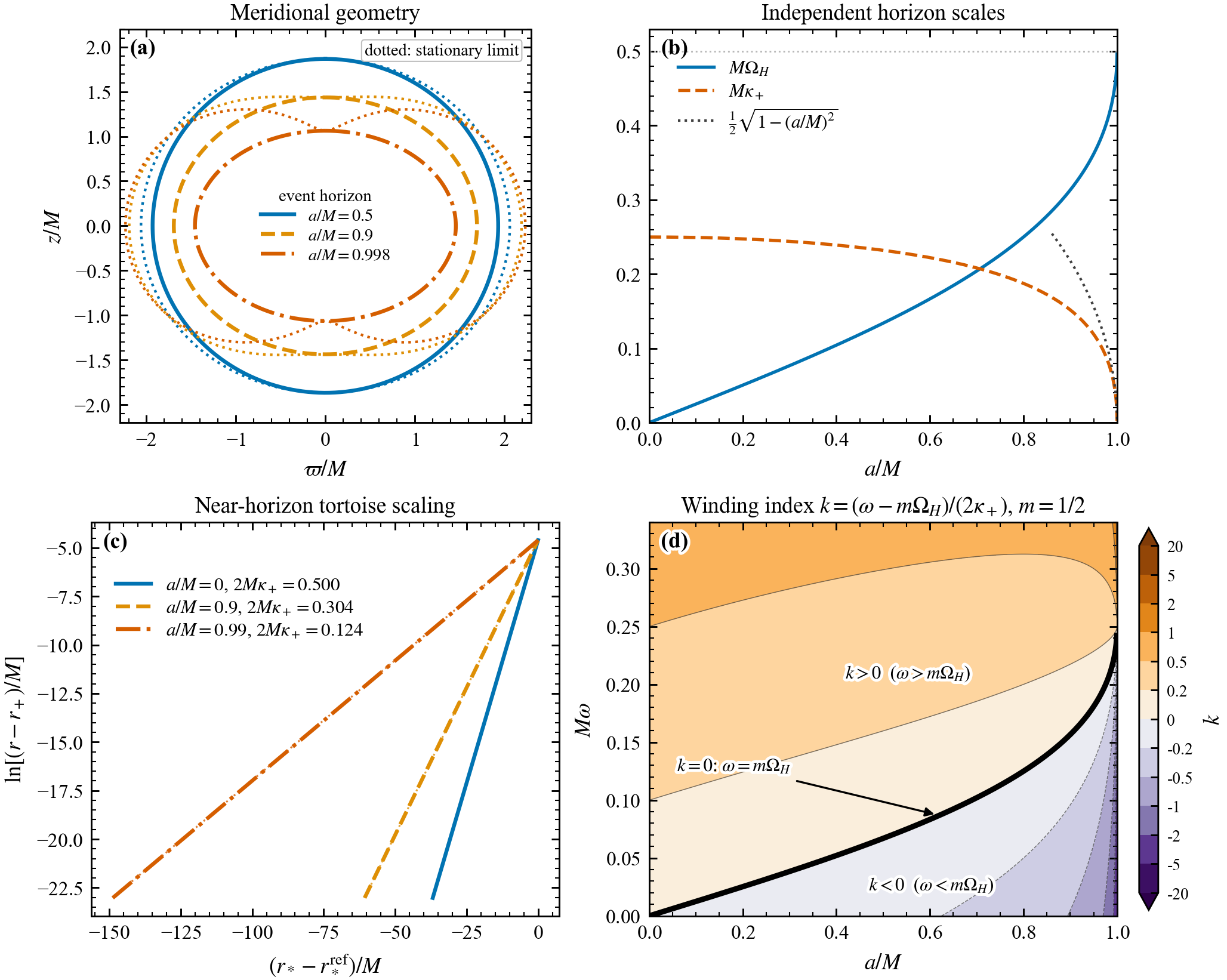}
\caption{Kerr geometry and the winding invariant $k$ (for $M=1$ and the fundamental Dirac mode $m=1/2$). (a) Meridional profiles of the event horizon (solid, dashed and dash-dotted) and stationary limit (dotted), drawn in oblate-spheroidal coordinates $\varpi=\sqrt{r^2+a^2}\sin\theta$, $z=r\cos\theta$. Increasing spin $a$ flattens the horizon and expands the ergoregion. (b) Horizon angular velocity $\Omega_H$ and surface gravity $\kappa_+$ versus spin (dotted: near-extremal asymptote $M\kappa_+\simeq\frac12\sqrt{1-a^2/M^2}$). The extremal limits ($\Omega_H \to 1/2M$, $\kappa_+ \to 0$) directly demonstrate their kinematic decoupling. (c) Near-horizon divergence of the tortoise coordinate $r_*$. Near the horizon $\ln[(r-r_+)/M]$ grows linearly with $r_*$, with slope $2\kappa_+$ [Eq.~\eqref{eq-2.9}]. (d) Parameter space of $k$ (orange: $k>0$; purple: $k<0$). The solid black curve represents the synchronization condition $\omega = m\Omega_H$, across which the winding invariant exactly vanishes and flips sign.
\label{fig:geometry}}
\end{figure*}

Unlike in spherical symmetry, $\Omega_{H}$ and $\kappa_{+}$ are kinematically independent in the Kerr geometry. Their Jacobian determinant
\begin{equation}
\det\frac{\partial(\Omega_{H},\kappa_{+})}{\partial(M,a)}
=\frac{1}{4M^{4}\,\delta\,(1+\delta)^{2}} ,
\qquad \delta\equiv\sqrt{1-a^{2}/M^{2}} ,
\label{eq-2.6}
\end{equation}
is strictly positive for $a<M$, diverging only at extremality ($a \to M$). However, the redundancy of $\kappa_+$ in the Schwarzschild limit is best understood via the dimensionless ratio
\begin{equation}
\frac{\Omega_{H}}{\kappa_{+}}=\frac{2a}{r_{+}-r_{-}}=\frac{a}{\sqrt{M^{2}-a^{2}}} .
\label{eq-2.7}
\end{equation}
At $a=0$, $\kappa_{+}r_{+} = 1/2$ is a geometric constant, rendering surface gravity parametrically redundant once $r_+$ sets the length scale. Conversely, as $a \to M$, $\kappa_{+} \to 0$ while $\Omega_{H} \to 1/(2M)$ (Fig.~\ref{fig:geometry}b). For near-horizon asymptotics, we introduce the tortoise coordinate $r_{*}$:
\begin{eqnarray}
&&\frac{dr_{*}}{dr}=\frac{r^{2}+a^{2}}{\Delta},  \nonumber \\
&&r_{*}=r+\frac{2Mr_{+}}{r_{+}-r_{-}}\ln\frac{r-r_{+}}{2M}
-\frac{2Mr_{-}}{r_{+}-r_{-}}\ln\frac{r-r_{-}}{2M} .
\label{eq-2.8}
\end{eqnarray}
Near the horizon, $dr_{*}/dr\simeq[2\kappa_{+}(r-r_{+})]^{-1}$, yielding
\begin{equation}
r-r_{+}\ \propto\ e^{2\kappa_{+}r_{*}} \qquad (r\to r_{+}) .
\label{eq-2.9}
\end{equation}
Thus, $r_{*} \to -\infty$ exponentially at a rate dictated solely by $\kappa_+$ (Fig.~\ref{fig:geometry}c). This logarithmic stretching breaks down completely at extremality ($a=M$), where $\Delta=(r-M)^{2}$ yields a leading-order pole:
\begin{equation}
r_{*}=r+2M\ln(r-r_{+})-\frac{2M^{2}}{r-r_{+}}+{\rm const}\qquad (a=M) ,
\label{eq-2.10}
\end{equation}
which alters the near-horizon phase structure.

\par
Separation of the Dirac equation utilizes the Newman--Penrose (NP) formalism. In the Kinnersley tetrad, aligned with the principal null congruences, the contravariant components are
\begin{eqnarray}
&l^{\mu}=\Big(\frac{r^{2}+a^{2}}{\Delta},\,1,\,0,\,\frac{a}{\Delta}\Big),~~n^{\mu}=\Big(\frac{r^{2}+a^{2}}{2\Sigma},\,-\frac{\Delta}{2\Sigma},\,0,\, \frac{a}{2\Sigma}\Big), \nonumber \\
&m^{\mu}=\frac{1}{\sqrt2\,\bar\rho} \Big(ia\sin\theta,\,0,\,1,\,\frac{i}{\sin\theta}\Big),~~~\bar\rho=r+ia\cos\theta ,
\label{eq-2.11}
\end{eqnarray}
satisfying the orthogonality relations
\begin{equation}
l\!\cdot\! l=n\!\cdot\! n=m\!\cdot\! m=0,\quad
l\!\cdot\! n=-1,\quad m\!\cdot\!\bar m=+1 .
\label{eq-2.12}
\end{equation}
The inverse metric consistency,
\begin{equation}
g^{\mu\nu}=-l^{\mu}n^{\nu}-n^{\mu}l^{\nu}+m^{\mu}\bar m^{\nu}+\bar m^{\mu}m^{\nu} ,
\label{eq-2.13}
\end{equation}
is utilized analytically and confirmed numerically (residuals $\lesssim 10^{-16}$) throughout our integration. The Kinnersley tetrad is singular on the future horizon ($l^{t,\varphi}\propto \Delta^{-1}$), and so is the Carter frame implicit in the variables $w_{1,2}$ of Sec.~\ref{sec:2.2}. The index $k$ defined below is frame independent, but statements about horizon regularity require a horizon-regular frame (Sec.~\ref{sec:3.2}).

\par
The separated spinor ansatz is
\begin{equation}
\psi\ \propto\ e^{i(m\varphi-\omega t)}\,R_{\mp1/2}(r)\,S_{\mp1/2}(\theta) ,
\label{eq-2.14}
\end{equation}
where $m$ is the half-integer azimuthal number, $\tilde m$ is the spinor mass, and $\lambda$ is the separation constant. In the Kinnersley tetrad the NP components of the spinor have moduli $|F_1|=|R_{-1/2}S_{-1/2}|/\sqrt{2\Sigma}$, $|F_2|=|R_{+1/2}S_{+1/2}|$, $|G_1|=|R_{+1/2}S_{-1/2}|$, and $|G_2|=|R_{-1/2}S_{+1/2}|/\sqrt{2\Sigma}$, where $F_1,G_2$ and $F_2,G_1$ carry boost weights $-1/2$ and $+1/2$~\cite{Chandrasekhar:1984siy}; the angular functions are normalized as $\int_0^\pi(S_{+1/2}^2+S_{-1/2}^2)\sin\theta\,d\theta=1$. Defining the differential operators
\begin{eqnarray}
&&\mathcal D_{n}=\partial_{r}-\frac{iK}{\Delta}+2n\frac{r-M}{\Delta},~~~K=(r^{2}+a^{2})\omega-am, \nonumber \\
&&\mathcal L_{n}=\partial_{\theta}+Q+n\cot\theta,~~~Q=-a\omega\sin\theta+\frac{m}{\sin\theta},
\label{eq-2.15}
\end{eqnarray}
with daggers $\mathcal D_{n}^{\dagger}$ and $\mathcal L_{n}^{\dagger}$ corresponding to $K\to-K$ and $Q\to-Q$, the Chandrasekhar--Page system reads:
\begin{eqnarray}
&&\Delta^{1/2}\mathcal D_{0}R_{-1/2}=(\lambda+i\tilde mr)\,\Delta^{1/2}R_{+1/2} ,
\label{eq-2.16} \\
&&\Delta^{1/2}\mathcal D_{0}^{\dagger}\big(\Delta^{1/2}R_{+1/2}\big)=(\lambda-i\tilde mr)\,R_{-1/2} ,
\label{eq-2.17}
\end{eqnarray}
\begin{eqnarray}
&&\mathcal L_{1/2}S_{+1/2}=-(\lambda-a\tilde m\cos\theta)\,S_{-1/2} ,
\label{eq-2.18}\\
&&\mathcal L_{1/2}^{\dagger}S_{-1/2}=+(\lambda+a\tilde m\cos\theta)\,S_{+1/2} .
\label{eq-2.19}
\end{eqnarray}

\par
With this sign of $iK/\Delta$, $\mathcal D_0$ is $l^\mu\partial_\mu$ acting on the mode~\eqref{eq-2.14}, and for $\tilde m=0$ Eqs.~\eqref{eq-2.16}--\eqref{eq-2.19} reduce to the Teukolsky equations for $s=\mp1/2$~\cite{Teukolsky:1973ha}, as we have checked symbolically. The same operators, with the same $\tilde m$, are used throughout. This system departs from the Schwarzschild case in three ways. First, the eigenvalue $\lambda$ depends implicitly on $\omega$, yielding a nonlinear two-parameter eigenvalue problem~\cite{Guo:2026pdi}, which we solve by root finding on the combined radial--angular residual. Second, the term $a\tilde m\cos\theta$ couples rotation linearly to the spinor mass, independently of $a\omega$; for a massive scalar field the mass enters the spheroidal problem only through $c^2=a^2(\omega^2-\mu^2)$~\cite{Teukolsky:1973ha,Berti:2005gp,Brito:2015rjv}. Third, the $-a\omega\sin\theta$ term promotes the angular basis to spin-weighted spheroidal harmonics. The Schwarzschild operators are trivially recovered as $a\to0$.

\par
At the horizon,
\begin{equation}
K(r_{+})=(r_{+}^{2}+a^{2})\big(\omega-m\Omega_{H}\big) .
\label{eq-2.20}
\end{equation}
Combining this with $\Delta'(r_{+})=2\kappa_{+}(r_{+}^{2}+a^{2})$ yields the simple pole residue of the radial operator:
\begin{equation}
\frac{iK}{\Delta}\ \xrightarrow{\ r\to r_{+}\ }\ \frac{ik}{r-r_{+}},
\qquad
k\equiv\frac{\omega-m\Omega_{H}}{2\kappa_{+}}=\frac{\omega-m\Omega_{H}}{4\pi T_{H}} .
\label{eq-2.21}
\end{equation}
Constructed from geometric invariants, the dimensionless winding number $k$ imprints the Hawking temperature onto the local wave dynamics. As $a\to0$, we recover the static-frame Frobenius index $k=\omega r_{+}$. To contextualize, for $M=1, m=1/2$: mapping $\omega=0.3$ yields $k\approx 0.472$ at $a=0.9$, and $k\approx 0.672$ at $a=0.99$ (Fig.~\ref{fig:geometry}d). Finally, incorporating an inhomogeneous source~\cite{Dzhunushaliev:2026rrt} into the radial system yields:
\begin{eqnarray}
&&\Delta^{1/2}\mathcal D_{0}R_{-1/2} =(\lambda+i\tilde mr)\Delta^{1/2}R_{+1/2} +\mathcal J_{-} ,
\label{eq-2.23}\\
&&\Delta^{1/2}\mathcal D_{0}^{\dagger}\big(\Delta^{1/2}R_{+1/2}\big)=(\lambda-i\tilde mr)R_{-1/2}+\mathcal J_{+} .
\label{eq-2.24}
\end{eqnarray}
To preserve separability, this source must inherit the angular projection of the spinor:
\begin{equation}
J\ \propto\ e^{i(m\varphi-\omega t)}\,\mathcal J_{\mp}(r)\,S_{\mp1/2}(\theta) .
\label{eq-2.25}
\end{equation}
Since the spheroidal harmonics depend on $\omega$, the angular source profile cannot be specified a priori. Its amplitude is constrained by Frobenius matching to the prescribed static-frame asymptotics; this matching does not impose smoothness on the future horizon.

\par
Equations~\eqref{eq-2.23}--\eqref{eq-2.24} are the separated form of the covariant sourced Dirac equation (in the conventions of Ref.~\cite{Dzhunushaliev:2026rrt})
\begin{equation}
\big(i\gamma^{\mu}\nabla_{\mu}-\tilde m\big)\Psi=\mathcal J ,
\label{eq-cov}
\end{equation}
where the NP components of the c-number source $\mathcal J$ are built from $\mathcal J_{\mp}(r)S_{\mp1/2}(\theta)$ in the same way as those of $\Psi$ are built from $R_{\mp1/2}(r)S_{\mp1/2}(\theta)$. Together with its Dirac adjoint, Eq.~\eqref{eq-cov} gives
\begin{equation}
\nabla_{\mu}j^{\mu}=\mathcal S\equiv 2\,{\rm Im}\big(\bar\Psi\mathcal J\big),\qquad j^{\mu}=\bar\Psi\gamma^{\mu}\Psi ,
\label{eq-cont}
\end{equation}
so the source creates ($\mathcal S>0$) or removes ($\mathcal S<0$) Dirac charge. For a stationary configuration the flux absorbed by the horizon therefore equals the flux entering from infinity plus the total creation rate; the separated form of this balance is derived in Sec.~\ref{sec:3.2}.

\subsection{Frobenius horizon asymptotics and the phase winding}
\label{sec:2.2}
\par
Near a subextremal horizon ($a<M$), we must disentangle the invariant kinematic detuning $\omega-m\Omega_H$ from coordinate artifacts introduced by Chandrasekhar's radial weights. The phase winding discussed below is a horizon effect; it does not arise for horizonless strongly gravitating spinor configurations, such as the Einstein--Dirac--Maxwell wormholes of Refs.~\cite{Konoplya:2021hsm,Dzhunushaliev:2025ntr,Dzhunushaliev:2025lki}. Defining
\begin{equation}
w_1=\Delta^{1/4}R_{+1/2},\qquad
w_2=\Delta^{-1/4}R_{-1/2} ,
\label{eq-2.26}
\end{equation}
and substituting into Eqs.~\eqref{eq-2.16}--\eqref{eq-2.17} yields the symmetric form
\begin{eqnarray}
&&w_2'+\left(\frac{\Delta'}{4\Delta}-\frac{iK}{\Delta}\right)w_2 =(\lambda+i\tilde m r)\Delta^{-1/2}w_1,
\label{eq-2.27}\\
&&w_1'+\left(\frac{\Delta'}{4\Delta}+\frac{iK}{\Delta}\right)w_1 =(\lambda-i\tilde m r)\Delta^{-1/2}w_2.
\label{eq-2.28}
\end{eqnarray}
In the presence of the source, Eqs.~\eqref{eq-2.23}--\eqref{eq-2.24}, the right-hand sides of Eqs.~\eqref{eq-2.27} and~\eqref{eq-2.28} acquire the terms $\mathcal J_2\equiv\Delta^{-3/4}\mathcal J_-$ and $\mathcal J_1\equiv\Delta^{-3/4}\mathcal J_+$, respectively. Equations~\eqref{eq-2.27}--\eqref{eq-2.28}, with these source terms, are the equations integrated numerically in this work (Appendix~\ref{app:repro}). Introducing the radial distance $\varepsilon\equiv r-r_+$ and the horizon separation $d\equiv r_+-r_->0$, the coefficients exhibit the limits
\begin{eqnarray}
\varepsilon\frac{\Delta'}{4\Delta} \longrightarrow\frac14,~~~\varepsilon\frac{iK}{\Delta} \longrightarrow ik,~~~\varepsilon\Delta^{-1/2}=\sqrt{\frac{\varepsilon}{d+\varepsilon}}\longrightarrow0.
\label{eq-2.29}
\end{eqnarray}
For $w=(w_1,w_2)^{\mathsf T}$, the homogeneous system $\varepsilon w'=A(\varepsilon) w$ has the leading matrix and eigenvalues:
\begin{equation}
A(0)=\operatorname{diag}\big[-\big(\tfrac14+ik\big),-\big(\tfrac14-ik\big)\big],~~~s_\pm=-\tfrac14\pm ik.
\label{eq-2.30}
\end{equation}
Because off-diagonal terms scale as $\varepsilon^{1/2}$, the Frobenius series expands naturally in $x=\sqrt{\varepsilon}$, rendering the matrix analytic~\cite{CoddingtonLevinson:1955,Wasow:1965}.

\par
\begin{figure*}[t]
\centering
\includegraphics[width=14cm]{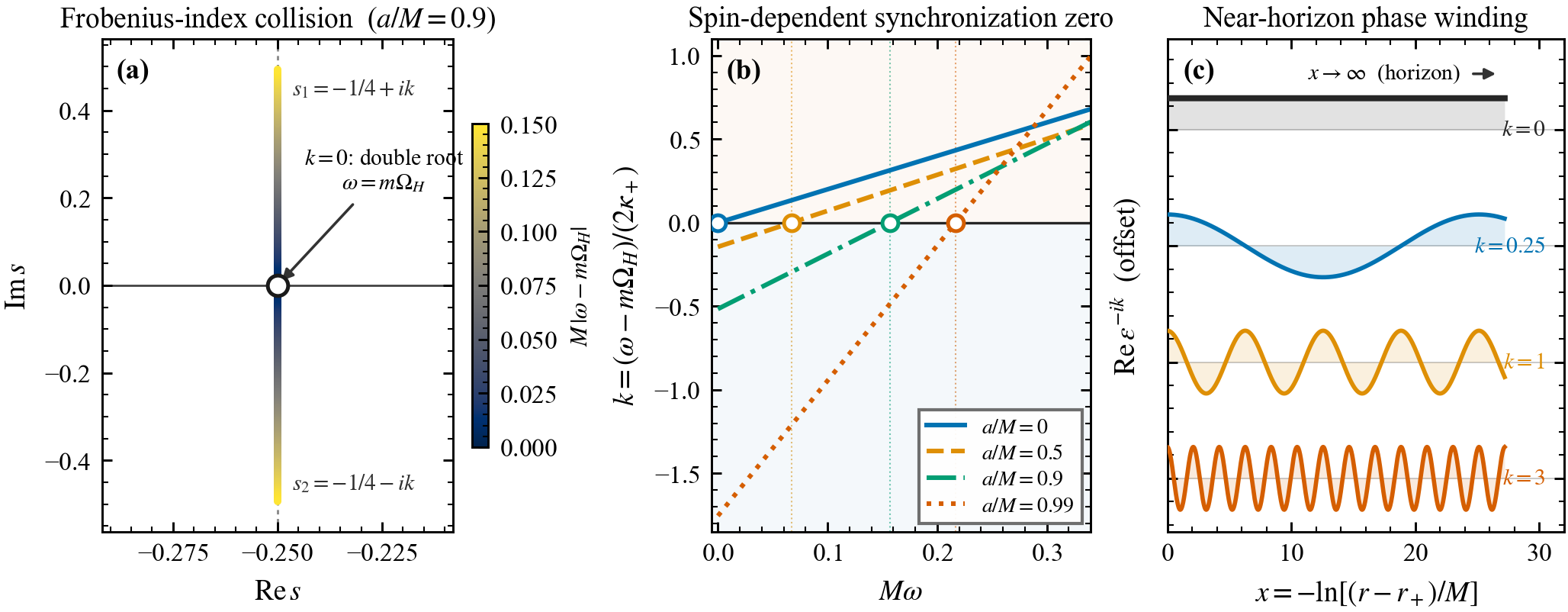}
\caption{Near-horizon indices and Boyer--Lindquist radial phase winding for $M=1$ and $m=\tfrac12$. (a) Trajectories of the two indices $s=-\tfrac14\pm ik$ in the complex plane at $a/M=0.9$, parameterized by the absolute detuning $|M\omega-m\Omega_H|$ (colormap). Both roots remain on $\operatorname{Re}s=-1/4$ and coalesce into a double root at synchronization. (b) Linear dependence of the winding parameter $k$ on $M\omega$ for various dimensionless spins; open circles mark $\omega=m\Omega_H$. (c) The real part of $(\varepsilon/M)^{-ik}$ as a function of the coordinate $x=-\ln[(r-r_+)/M]$, with $\varepsilon\equiv r-r_+$ (curves are vertically offset for clarity). The radial winding becomes faster as $|k|$ increases and disappears at $k=0$. This panel illustrates the Boyer--Lindquist radial factor, rather than a covariant regularity test.}
\label{fig:indices}
\end{figure*}

Transforming back to original variables, the purely local radial branches scale as
\begin{eqnarray}
&&R_{-1/2}^{(1)} \sim(\varepsilon/M)^{+ik},~~\Delta^{1/2}R_{+1/2}^{(1)}=O\!\left[(\varepsilon/M)^{1/2+ik}\right],\nonumber \\
&&R_{-1/2}^{(2)}=O\!\left[(\varepsilon/M)^{1/2-ik}\right],~~\Delta^{1/2}R_{+1/2}^{(2)}\sim(\varepsilon/M)^{-ik}.
\label{eq-2.31}
\end{eqnarray}
Combined with the mode factor, branch (1) is outgoing and branch (2) is ingoing at the horizon, in agreement with the Teukolsky asymptotics $R_s\sim\Delta^{-s}e^{-i(\omega-m\Omega_H)r_*}$ (ingoing) and $R_s\sim e^{+i(\omega-m\Omega_H)r_*}$ (outgoing) for $s=\pm1/2$~\cite{Teukolsky:1973ha,Teukolsky:1974yv}. The real part ($-1/4$) offsets the $\Delta$-weights, ensuring a finite-modulus leading amplitude. The imaginary part $ik$, however, induces a radial phase. Utilizing Eq.~\eqref{eq-2.9},
\begin{equation}
\left(\frac{\varepsilon}{M}\right)^{\pm ik}
=C_\pm e^{\pm i(\omega-m\Omega_H)r_*}\,[1+o(1)],
\qquad |C_\pm|=1.
\label{eq-2.32}
\end{equation}
Consequently, the phase winds indefinitely as $r_*\to-\infty$ for $k\neq0$; over an interval $L$ in $r_*$ it accumulates
\begin{equation}
N_{\rm wind}=\frac{|\omega-m\Omega_H|\,L}{2\pi}.
\label{eq-2.33}
\end{equation}
This Boyer--Lindquist phase winding is removable for the ingoing branch. In regular ingoing Kerr coordinates ($v=t+r_*$, $\widetilde\varphi=\varphi+\int a\,dr/\Delta$), the Fourier mode scales as $e^{i(m\widetilde\varphi-\omega v)} e^{+i(\omega-m\Omega_H)r_*}$. One radial branch in Eq.~\eqref{eq-2.32} cancels this divergent phase, yielding a smooth ingoing wave. For the outgoing branch the phase doubles, $e^{2i(\omega-m\Omega_H)r_*}$, and in a horizon-regular frame its boost-weight $-1/2$ component diverges as $\varepsilon^{-1/2}$ (Sec.~\ref{sec:3.2}), so regularity on the future horizon selects the ingoing solution. Thus, stationary source-free solutions are locally permissible, as a one-parameter family; their global preclusion stems from the flux identity derived in Sec.~\ref{sec:3.2}. This winding vanishes at exact synchronization, $\omega=m\Omega_H$~\cite{Hod:2012px,Brito:2015rjv,Teukolsky:1974yv} (see Fig.~\ref{fig:indices}).

\par
The index collision at $k=0$ generates no logarithmic branch. Writing $\widetilde A(x)=\sum_{j}\widetilde A_jx^j$ and $w=x^{\tilde s}\sum_{n} a_nx^n$ for $x\,d w/dx=\widetilde A(x) w$, we find:
\begin{eqnarray}
&&\widetilde{A}_0  = \operatorname{diag}\left(-\frac{1}{2}-2ik, -\frac{1}{2}+2ik\right), \nonumber \\
&&\tilde{s}_\pm  = -\frac{1}{2} \pm 2ik, \nonumber \\
&&[(\tilde{s}_\pm + n) \mathbf{I}_2 - \widetilde{A}_0] a_n  = \sum_{j=1}^{n} \widetilde{A}_j a_{n-j}, \nonumber \\
&&\det[(\tilde{s}_\pm + n) \mathbf{I}_2 - \widetilde{A}_0]  = n(n\pm 4ik).
\label{eq-2.34}
\end{eqnarray}
For real $k$ and $n\geq 1$, the determinant is nonzero (it is complex for $k\neq0$, with modulus $n\sqrt{n^2+16k^2}\geq n^2$), which precludes any Frobenius resonance. At synchronization ($k=0$), $\widetilde{A}_0=-\tfrac{1}{2}\mathbf{I}_2$; the fundamental eigenvector $a_0$ remains arbitrary, and all $a_n$ are determined algebraically without logarithms; regularity on the future horizon nevertheless restricts it to $a_0\propto(1,0)^{\mathsf T}$, as shown in Sec.~\ref{sec:3.2}. (Imposing a nonoscillatory condition $w_i \sim b_i\varepsilon^p$ forces the source amplitudes to $\mathcal J_{1,2}$ $\sim \varepsilon^{p-1}$, though this does not uniquely exclude underlying homogeneous modes.)

\par
Although both local solutions at $k=0$ are free of logarithms and scale as $\varepsilon^{-1/4}$ in $w_{1,2}$, regularity still selects one of them: in a horizon-regular frame (Sec.~\ref{sec:3.2}) the ingoing solution, $a_0\propto(1,0)^{\mathsf T}$, is smooth, whereas the boost-weight $-1/2$ component of the outgoing one, $a_0\propto(0,1)^{\mathsf T}$, diverges as $\varepsilon^{-1/2}$ (Sec.~\ref{sec:4.2}). For scalar modes the excluded solution is logarithmic instead~\cite{CoddingtonLevinson:1955}. In both cases the existence of bound states is decided by the flux of the selected solution (Sec.~\ref{sec:3.2}).

\par
\begin{figure*}[t]
\centering
\includegraphics[width=14cm]{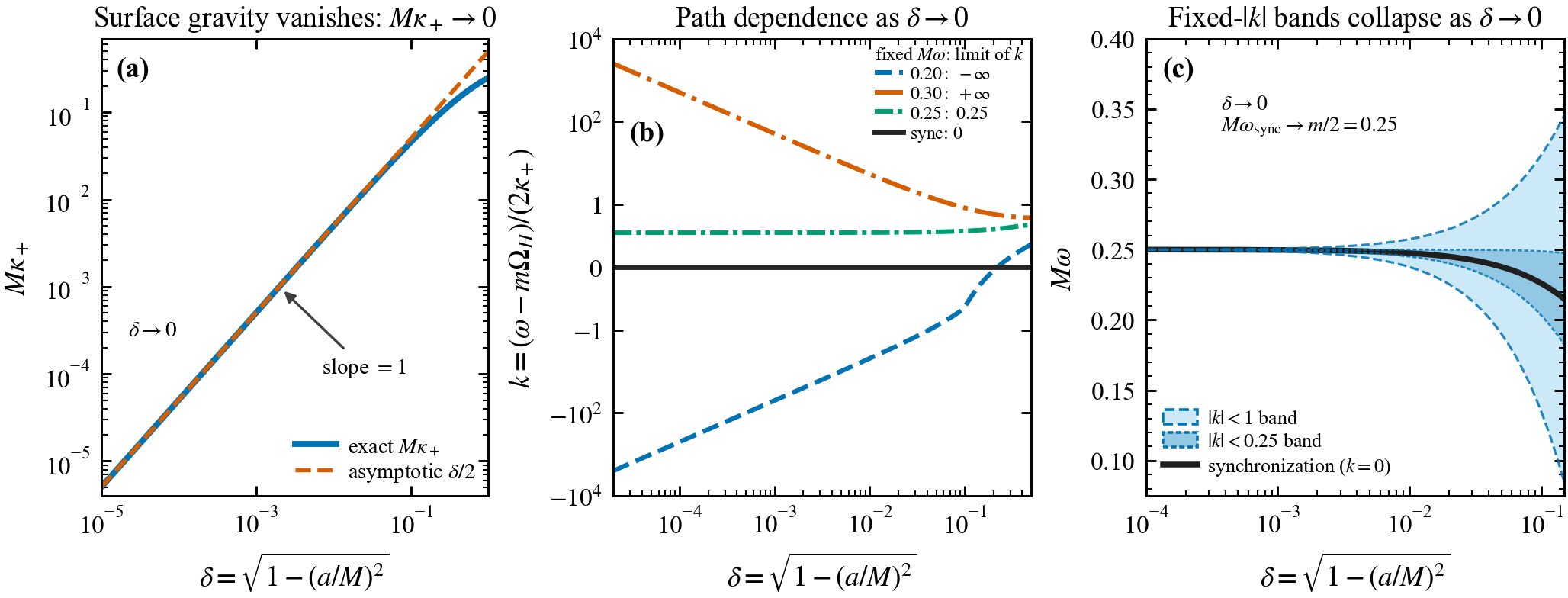}
\caption{Path dependence of the near-extremal limit for $M=1$ and $m=\tfrac12$. (a) The surface gravity vanishes linearly, with the asymptotic scaling $M\kappa_+ \sim \delta/2$ where $\delta=\sqrt{1-(a/M)^2}$. (b) The winding invariant $k$ along fixed-frequency paths. As $\delta \to 0$, $k$ diverges for generic detuned frequencies, approaches a finite value along the critical path $M\omega=m/2$, and remains exactly zero at synchronization. (c) The collapse of frequency bands. Any region bounded by a finite $|k|$ physically narrows proportionally to $\kappa_+$, collapsing entirely onto the synchronization limit $M\omega=m/2$ at extremality.}
\label{fig:extremal}
\end{figure*}

In the near-extremal regime ($\delta\equiv\sqrt{1-a^2/M^2} \to 0$), the kinematics scale as
\begin{equation}
M\Omega_H=\frac12-\frac{\delta}{2}+O(\delta^2), \qquad M\kappa_+=\frac{\delta}{2}+O(\delta^2).
\label{eq-2.35}
\end{equation}
For a fixed frequency $\omega$, the invariant index $k$ diverges:
\begin{eqnarray}
k && =\frac{(1+\delta)M\omega-\dfrac{m}{2}\sqrt{1-\delta^2}}{\delta} \nonumber \\
&&=\frac{M\omega-\dfrac{m}{2}}{\delta}+M\omega+\frac{m}{4}\delta+O(\delta^3).
\label{eq-2.36}
\end{eqnarray}
To maintain a bounded $k$, the kinematic detuning must vanish at $\mathcal{O}(\delta)$. Setting $M\omega=\frac{m}{2}+\alpha\delta$ gives $k=\frac{m}{2}+\alpha$. Thus, while bounded solutions fix the limiting frequency at $m/(2M)$, the resulting index remains strongly path-dependent: a fixed-frequency trajectory yields $k \to m/2$, whereas exact synchronization ($\alpha=-m/2$) enforces $k\equiv 0$. Non-extremal boundary conditions therefore cannot be naively extrapolated across extremality (Fig.~\ref{fig:extremal}).

\par
In the extremal Kerr--Newman spacetime ($M^2=a^2+Q^2$, $r_+=M$),
\begin{equation}
\Omega_H=\frac{a}{M^2+a^2},\qquad \Phi_H=\frac{QM}{M^2+a^2}.
\label{eq-2.37}
\end{equation}
Schmid's extremal condition $\omega(M^2+a^2)+k_{\rm S}a+eQM=0$~\cite{Schmid:2002zf} reduces in our gauge ($m=-k_{\rm S}$, $q=-e$) to $\omega=m\Omega_H+q\Phi_H$. While marking the superradiant threshold for charged bosons, for fermions it acts as a horizon-resonance condition~\cite{Brito:2015rjv}. Uncharged, this recovers $\omega=m/(2M)$, a necessary but not sufficient condition for extremal Dirac bound states~\cite{Schmid:2002zf}; for bosonic perturbations the same limit governs the long-lived (zero-damped) quasinormal modes of nearly extremal Kerr black holes~\cite{Hod:2008zz,Yang:2013uba}.

\par
Ultimately, the exact extremal limit ($a=M$) is governed by $K_H\equiv K(M)=2M^2[\omega-m/(2M)]$. For asynchronous modes ($K_H\neq0$), $K/\Delta\sim K_H/\varepsilon^2$ turns the horizon into an irregular singularity dominated by the essential phase $e^{\pm iK_H/\varepsilon}$. Conversely, exact synchronization ($K_H=0$) yields $K=m\varepsilon+(m/2M)\varepsilon^2$, canceling the second-order pole and reducing Eqs.~\eqref{eq-2.27}--\eqref{eq-2.28} to a Fuchsian system, i.e., the horizon becomes a regular singular point:
\begin{eqnarray}
&&\varepsilon w'=A_{\rm ext} w+O(\varepsilon) w, \nonumber \\
&&A_{\rm ext}=
\begin{pmatrix}
-\dfrac12-im & \lambda-i\tilde m M \\
\lambda+i\tilde m M & -\dfrac12+im
\end{pmatrix},\label{eq-2.38}\\
&&s_{\rm ext}=-\frac12\pm \nu,\qquad \nu\equiv \sqrt{\lambda^2+\tilde m^2M^2-m^2}. \nonumber
\end{eqnarray}
This non-diagonal leading matrix $A_{\rm ext}$ is distinct from the $k=0$ limit of Eq.~\eqref{eq-2.34}, a discrete algebraic jump that underlies the extremal bound states reported in Ref.~\cite{Schmid:2002zf}. According to Schmid, such normalizable, time-periodic solutions exist if and only if, besides $\omega=m/(2M)$ and $|\omega|<\tilde m$, one has $\lambda^2+a^2\tilde m^2-m^2>1/4$ (i.e., $\nu>1/2$) and a quantization condition holds; this requires $|m|/2<M\tilde m<|m|/\sqrt2$ and selects discrete values of $M\tilde m$ in this window~\cite{Schmid:2002zf} (see Sec.~\ref{sec:4.2} for a later objection). The $a \to M$ limit is therefore non-uniform.

\subsection{Physical origin of the source and status of its ingredients}
\label{sec:2.3}
\par
In Ref.~\cite{Dzhunushaliev:2026rrt} the source is introduced ad hoc, as a phenomenological representation of a hypothesized near-horizon modification of the canonical anticommutation relations; neither Ref.~\cite{Dzhunushaliev:2026rrt} nor the present work derives it from a deformed field algebra, so its physical origin remains an assumption. Linear, externally prescribed sources do arise in other settings. If the Dirac field mixes with another sector, e.g., through a Yukawa-type term $-g\phi\bar\psi\chi+{\rm h.c.}$ with a scalar background $\phi$ and a second fermion $\chi$, its field equation acquires the source $\mathcal J=g\phi\chi$. A stationary, co-rotating product, such as a scalar cloud with $(\omega_\phi,m_\phi)$ times a fermionic mode with $(\omega_\chi,m_\chi)$ (by Sec.~\ref{sec:3.2} this cannot be a normalizable, source-free bound state, but it can be a stationary scattering state maintained from infinity), has the time and azimuthal dependence of Eq.~\eqref{eq-2.25} with $\omega=\omega_\phi+\omega_\chi$ and $m=m_\phi+m_\chi$, and separability further requires an angular profile proportional to $S_{\mp1/2}(\theta)$. Since $\phi\chi$ is Grassmann odd in a quantized setting, the c-number treatment corresponds to the single-particle level or to the equation for a mixed correlation function. Inhomogeneous Dirac equations also arise for two-point functions with one argument held fixed; for the Feynman propagator the source is a $\delta$ function at coincident points~\cite{Peskin:1995ev}, and Ref.~\cite{Dzhunushaliev:2026rrt} postulates that near a horizon it becomes extended. The source is then a property of the state $|Q\rangle$ and of the field algebra, which would have to be specified to give the solutions a quantum meaning.

\par
The source can be treated as external if the sector producing it is not affected by the Dirac configuration (weak coupling, or a reservoir whose occupation greatly exceeds that of the bound state), if it is stationary and co-rotating with the same $(\omega,m)$ and angular structure as the field, and if the energy, angular momentum, and charge of the Dirac configuration and of the source sector are small compared with $M$, $J=aM$, and $Q$. Because the net horizon flux and the sphere-integrated creation rate vanish for all configurations considered here (Sec.~\ref{sec:3.2}), the source need not supply charge at a net rate. It is, however, singular at the horizon in the static frame [Eq.~\eqref{eq-4.8}], and whether a microscopic sector can produce such a profile is an open question.

\par
The source is not unique. For prescribed horizon data, local Frobenius matching (Sec.~\ref{sec:4.2}) fixes only its singular coefficients, Eq.~\eqref{eq-4.8}, in terms of the leading static-frame coefficient $w^{(0)}$ and $k$, and the next coefficient through the requirement, adopted from Ref.~\cite{Dzhunushaliev:2026rrt}, that the solution contain no $O(\varepsilon^{1/2})$ term in Case A [no $O(\varepsilon)$ term in Case B]. Postulated, mostly following Ref.~\cite{Dzhunushaliev:2026rrt}, are the existence and interpretation of the source, its separable angular structure~\eqref{eq-2.25}, its radial envelope $e^{-\Sigma(r-r_+)}$ with $\Sigma=\sqrt{\tilde m^2-\omega^2}$ (not the metric function $\Sigma$) and the prefactor $r/r_+$ in Eq.~\eqref{eq-4.12}, the horizon behaviors $w\propto\varepsilon^{p}$ with $p=0$ (Case A) or $p=1/2$ (Case B), and the amplitude ratio $v_0/u_0$. Derived from the field equations are the singular source coefficients, the index $k$ and the synchronization locus, the Frobenius structure (including the absence of logarithms at $k=0$), the regularity and flux properties of Sec.~\ref{sec:3.2}, and the angular spectrum. The eigenfrequencies and the branches $\omega(a)$ follow once all postulated elements are fixed; their dependence on these elements is studied in Sec.~\ref{sec:4.3}.

\section{Source matching, horizon regularity, and the flux identity}
\label{sec:3}
\par
\subsection{Source matching in the spherically symmetric limit}
\label{sec:3.1}
\par
To clarify near-horizon source matching and establish a numerical benchmark for $a\to0$, we temporarily adopt the notation of Ref.~\cite{Dzhunushaliev:2026rrt}. In terms of the dimensionless mass $m=r_H\tilde m$, frequency $\Omega=r_H\omega$, decay constant $\Sigma=\sqrt{m^2-\Omega^2}$, and radial coordinate $x=r/r_H-1$, the spherically symmetric Dirac radial equations for $\kappa=-1$ read~[Eqs.~(11)--(12) of Ref.~\cite{Dzhunushaliev:2026rrt}; see also Ref.~\cite{Chandrasekhar:1976ap}]:
\begin{eqnarray}
&&\sqrt f\,v'+\frac{1+\big(1+\tfrac{1}{4x}\big)\sqrt f}{x+1}\,v
+\Big(m-\frac{\Omega}{\sqrt f}\Big)u = j_1,
\label{eq-3.1}\\
&&\sqrt f\,u'+\frac{-1+\big(1+\tfrac{1}{4x}\big)\sqrt f}{x+1}\,u
+\Big(m+\frac{\Omega}{\sqrt f}\Big)v = j_2,
\label{eq-3.2}
\end{eqnarray}
where $f=x/(x+1)$ and primes denote $d/dx$. Near the horizon ($x\to 0$), the spin connection and frequency terms scale as $x^{-1/2}$, rendering the static-frame equations locally singular. The required source structure is thus fixed by the prescribed near-horizon asymptotics of the static-frame radial amplitudes $u$ and $v$, rather than by smoothness in a horizon-regular frame.

\par
Reference~\cite{Dzhunushaliev:2026rrt} explores two phenomenological source configurations; Eqs.~\eqref{eq-3.3}--\eqref{eq-3.5} below reproduce its Eqs.~(13)--(14), (17), (18)--(19), and (22), which we have rederived symbolically. Case A assumes finite static-frame limits, $v=v_0+\mathcal{O}(x)$ and $u=u_0+\mathcal{O}(x)$ as $x\to0^+$, supported by the profile:
\begin{eqnarray}
j_1 &=& e^{-\Sigma x}\left(\frac{q_0}{\sqrt{x}}+q_1\right),\nonumber\\
j_2 &=& e^{-\Sigma x}\left(\frac{p_0}{\sqrt{x}}+p_1\right),\nonumber\\
\Sigma &=& \sqrt{m^2-\Omega^2}.
\label{eq-3.3}
\end{eqnarray}
For this ansatz, matching the $x^{-1/2}$ and $x^0$ terms fixes the coefficients:
\begin{eqnarray}
q_0 &=& \tfrac14(v_0-4\Omega u_0), \qquad q_1 = mu_0+v_0,\nonumber\\
p_0 &=& \tfrac14(u_0+4\Omega v_0), \qquad p_1 = mv_0-u_0.
\label{eq-3.4}
\end{eqnarray}
Case B assumes static-frame radial amplitudes vanishing as $v=v_0\sqrt{x}+\cdots$ and $u=u_0\sqrt{x}+\cdots$. Assuming a source $j_1=e^{-\Sigma x}(q_1+q_2\sqrt{x})$ and $j_2=e^{-\Sigma x}(p_1+p_2\sqrt{x})$, order-by-order matching yields:
\begin{eqnarray}
q_1 &=& \tfrac14(3v_0-4\Omega u_0), \qquad q_2 = mu_0+v_0,\nonumber\\
p_1 &=& \tfrac14(3u_0+4\Omega v_0), \qquad p_2 = mv_0-u_0.
\label{eq-3.5}
\end{eqnarray}

\par
Because the system mixes integer and half-integer powers of $x$, the prescribed static-frame asymptotics are implemented by a Frobenius expansion in $t=\sqrt{x}$. Defining $\boldsymbol y=(v,u)^{\mathsf T}$, algebraic matching of the singular terms trivializes the linear $\mathcal{O}(t)$ coefficient in Case A and the quadratic $\mathcal{O}(t^2)$ coefficient in Case B, yielding the local series:
\begin{eqnarray*}
\boldsymbol y_{\mathrm A} &=& \boldsymbol y_0+\boldsymbol y_{\mathrm{A},2}x +\boldsymbol y_{\mathrm{A},3}x^{3/2}+\mathcal{O}(x^2),\\
\boldsymbol y_{\mathrm B} &=& \boldsymbol y_0x^{1/2}+\boldsymbol y_{\mathrm{B},2}x^{3/2} +\boldsymbol y_{\mathrm{B},3}x^2+\mathcal{O}(x^{5/2}).
\end{eqnarray*}
As an independent structural check, setting $j_1=j_2=0$ and transforming to $w_2 = (x+1)^{1/2}(v-iu)$ [equivalently $w_1=\overline{w_2}=(x+1)^{1/2}(v+iu)$, with $\lambda=-1$ and $\tilde m=m/r_H$] exactly recovers the sourceless Chandrasekhar--Page system [Eqs.~\eqref{eq-2.16} and~\eqref{eq-2.17}]~\cite{Chandrasekhar:1976ap}.

\par
The static-frame $x^{-1/2}$ divergence has a clear geometric origin:
\begin{equation}
\frac{1}{4x}\frac{\sqrt f}{x+1}
=\frac{\sqrt f}{4}\frac{d\ln f}{dx}
=\frac{1}{4\sqrt{x}(x+1)^{3/2}}.
\label{eq-3.6}
\end{equation}
Since $\sqrt f\,v'+\tfrac14\sqrt f\,(\ln f)'\,v=f^{1/4}\big(f^{1/4}v\big)'$, defining $\hat v=f^{1/4}v$ and $\hat u=f^{1/4}u$ absorbs this spin-connection artifact~\cite{Chandrasekhar:1976ap}. Physically, this singularity reflects the infinite proper acceleration $|a|$ required for a static observer to hover at the horizon~\cite{Wald:1984rg,Misner:1973prb}:
\begin{eqnarray}
a^{\mu} &=& u_{\mathrm{stat}}^{\nu}\nabla_{\nu}u_{\mathrm{stat}}^{\mu}= \left(0,\frac{M}{r^2},0,0\right),\nonumber\\
a_{\mu}a^{\mu} &=& \frac{M^2}{r^4f} \implies |a| = \frac{M}{r^2\sqrt f}\longrightarrow\infty.
\label{eq-3.7}
\end{eqnarray}
This singular frame coefficient reflects a kinematic pathology of the observer congruence; by itself it does not determine whether the spinor is singular in a horizon-regular frame (Sec.~\ref{sec:3.2}). Under this rescaling, the homogeneous equations yield indices $\hat s=\pm i\Omega$. With $r_*\simeq r_H\ln x$, this recovers the standard phase winding $x^{\pm i\Omega}\simeq e^{\pm i\omega r_*}$~\cite{Chandrasekhar:1976ap,Chandrasekhar:1984siy}. Transformed to advanced Eddington--Finkelstein coordinates, the ingoing phase is regular. Consequently, the local static-frame breakdown does not constitute a non-existence theorem; it simply precludes the non-oscillatory power series assumed in Cases A and B for any $\Omega \neq 0$. The exclusion of source-free stationary bound states is global~\cite{Finster:1998ak,Finster:1999ry}; we formulate it within the separated framework in Sec.~\ref{sec:3.2}.

\par
\begin{figure*}[t]
\centering
\includegraphics[width=14cm]{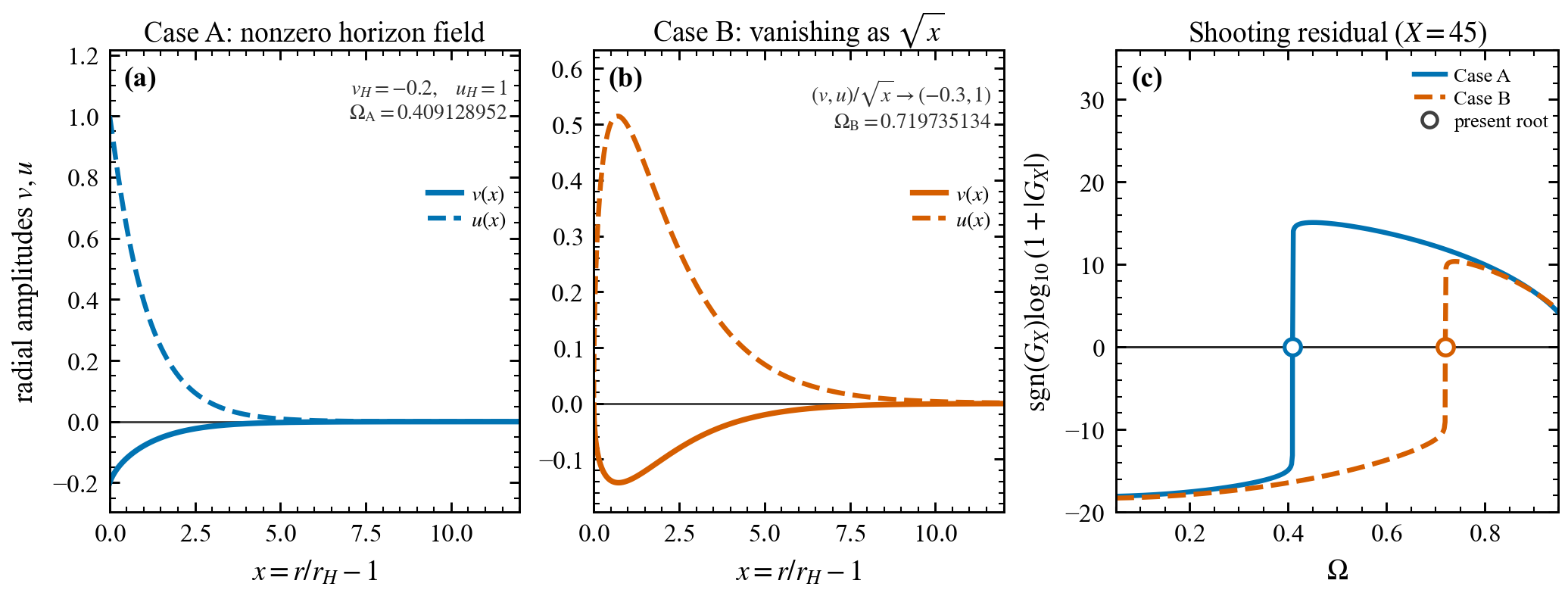}
\caption{Spherically symmetric benchmark for $m=1$ evaluated at a matching radius $X=45$. (a) Radial amplitudes for Case A initialized with $(v_0,u_0)=(-0.2,1)$, prescribing nonzero limits of the static-frame radial amplitudes. (b) Radial amplitudes for Case B initialized with $(v_0,u_0)=(-0.3,1)$, where the static-frame radial amplitudes vanish as $\sqrt{x}$ as $x\to0^+$. (c) The mapped shooting residual $\operatorname{sgn}(G_X)\log_{10}(1+|G_X|)$ across a sweep of frequencies. Open circles mark the numerical roots given in Eq.~\eqref{eq-3.9}, whose sensitivity is discussed in Appendix~\ref{app:repro}.}
\label{fig:anchor}
\end{figure*}

To isolate true bound states, we enforce exponential decay as $x\to\infty$. Evaluating the numerical solution at a large matching radius $X$, we construct the projection operator:
\begin{equation}
G_X(\Omega)\equiv
\left.\left(\frac{v}{m-\Omega}-\frac{u}{\Sigma}\right)\right|_{x=X}.
\label{eq-3.8}
\end{equation}
Imposing $G_X(\Omega)=0$ enforces the leading asymptotic condition for eliminating the growing mode at a finite matching radius; the growing mode has asymptotic direction $(v,u)\propto(m-\Omega,-\Sigma)$. The finite-radius approximation is checked by varying $X$. Integrating from a horizon expansion up to $\mathcal{O}(t^5)$ via a high-precision shooting algorithm, we find the roots for $m=1$ with initial data $(v_0,u_0)=(-0.2,1)$ (Case A) and $(-0.3,1)$ (Case B) (Fig.~\ref{fig:anchor}):
\begin{eqnarray}
\Omega_{\mathrm A} &=& 0.409128952,\nonumber\\
\Omega_{\mathrm B} &=& 0.719735134.
\label{eq-3.9}
\end{eqnarray}
While Ref.~\cite{Dzhunushaliev:2026rrt} (Fig.~1 there) reported $\Omega_{\mathrm A}=0.409078$ and $\Omega_{\mathrm B}=0.7197343$, the tests in Appendix~\ref{app:repro} give maximum observed changes of $10^{-11}$ in $\Omega_{\mathrm A}$ and $4\times10^{-8}$ in $\Omega_{\mathrm B}$, using $\Omega=2M\omega$. The larger Case B change occurs at the smallest outer boundary, $R_{\rm out}=30M$; all other listed Case B changes are below $2\times10^{-11}$ in $\Omega$. Here $X=R_{\rm out}/r_H$, so $R_{\rm out}=30M$ corresponds to $X=15$. These variations measure numerical sensitivity and are not certified error bounds. The same horizon data also admit negative-frequency roots, $\Omega=-0.4472694$ (Case A) and $\Omega=-0.6431728$ (Case B), outside the range considered in Ref.~\cite{Dzhunushaliev:2026rrt}; they reappear in the Kerr branch structure of Sec.~\ref{sec:4.2}. These numerical roots serve as benchmarks for our Kerr code in the limit $a\to0$, with the accuracy qualifications in Appendix~\ref{app:repro}.

\subsection{Frames, horizon regularity, and the flux identity}
\label{sec:3.2}
\par
Beyond the local matching of Sec.~\ref{sec:3.1}, the role of the source is clarified by the causal structure of the Dirac current. For any spinor $\psi$, the number current $j^{\mu}=\bar\psi\gamma^{\mu}\psi$ is causal. The time component in any orthonormal frame is positive for non-vanishing fields:
\begin{equation}
j^{0}=\psi^{\dagger}\psi>0 \qquad (\psi\neq0) .
\label{eq-3.10}
\end{equation}
Furthermore, adopting the $(-,+,+,+)$ metric signature, the Fierz identity gives the causal bound
\begin{equation}
j^{a}j_{a}=-\big(S^{2}+P^{2}\big)\le0 , \qquad S=\bar\psi\psi , \quad P=i\bar\psi\gamma^{5}\psi .
\label{eq-3.11}
\end{equation}
Combined, Eqs.~\eqref{eq-3.10} and~\eqref{eq-3.11} confine $j^{\mu}$ on or within the future light cone. The current is null if and only if $S=P=0$ (e.g., for a pure chirality state), in which case it is still future-directed. As a covariant vector relation, this causal structure is frame independent.

\par
Three frames appear in this work. The Kinnersley tetrad, Eq.~\eqref{eq-2.11}, used for the separation, is singular on the future horizon $\mathcal H^+$. The Carter orthonormal frame
\begin{eqnarray}
&&e_0=\frac{(r^2+a^2)\partial_t+a\partial_\varphi}{\sqrt{\Sigma\Delta}},\qquad e_1=\sqrt{\frac{\Delta}{\Sigma}}\,\partial_r,\nonumber\\
&&e_2=\frac{\partial_\theta}{\sqrt\Sigma},\qquad e_3=\frac{\partial_\varphi+a\sin^2\theta\,\partial_t}{\sqrt\Sigma\,\sin\theta},
\label{eq-carter}
\end{eqnarray}
is adapted to the Carter observers $e_0$, which reduce to the static observers for $a=0$ and become null at the horizon. It differs from the Kinnersley tetrad by a boost and a spin rotation, and up to a common factor $\Sigma^{-1/4}$ and phases, $w_{1,2}$ and the combinations $v\pm iu$ of Sec.~\ref{sec:3.1} are its components; ``static-frame regularity'' in Ref.~\cite{Dzhunushaliev:2026rrt} and in Sec.~\ref{sec:3.1} means that these components are finite. A horizon-regular frame is obtained from the Kinnersley tetrad by the boost~\cite{Hawking:1972hy,Teukolsky:1974yv}
\begin{equation}
l'=A\,l,\quad n'=A^{-1}n,\quad m'=m,\quad A=\frac{\Delta}{2(r^2+a^2)} .
\label{eq-boost}
\end{equation}
In ingoing Kerr coordinates $(v,r,\theta,\widetilde\varphi)$~\cite{Carter:1968rr}, $l'=\partial_v+\frac{\Delta}{2(r^2+a^2)}\partial_r+\frac{a}{r^2+a^2}\partial_{\widetilde\varphi}$ and $n'=-\frac{r^2+a^2}{\Sigma}\partial_r$ are smooth across $\mathcal H^+$, and $l'|_{\mathcal H}=\xi$ [Eq.~\eqref{eq-3.12} below]. Components of boost weight $\pm1/2$ are multiplied by $A^{\pm1/2}$~\cite{Geroch:1973am}, and $e^{i(m\varphi-\omega t)}=e^{i(m\widetilde\varphi-\omega v)}\varepsilon^{ik}[1+O(\varepsilon)]$, so that near $\mathcal H^+$ the field in this frame is governed by the radial factors
\begin{eqnarray}
&&Y_+\equiv\varepsilon^{ik}\Delta^{1/4}w_1=\varepsilon^{ik}\Delta^{1/2}R_{+1/2},\nonumber\\
&&Y_-\equiv\varepsilon^{ik}\Delta^{-1/4}w_2=\varepsilon^{ik}\Delta^{-1/2}R_{-1/2},
\label{eq-Y}
\end{eqnarray}
which multiply smooth angular factors; a configuration is regular on $\mathcal H^+$ if $Y_\pm$ are smooth at $r_+$. We use three invariant quantities: the flux density $-j_\mu\xi^\mu$ across $\mathcal H^+$; the density $n_{\rm ff}=-j_\mu u^\mu_{\rm ff}$ measured by freely falling observers, whose four-velocity has finite, positive components along $l'$ and $n'$ at $\mathcal H^+$; and the density $n_C=-j_\mu e_0^\mu$ measured by the Carter observers (Sec.~\ref{sec:4.1}). Note that $\sqrt{-g}=\Sigma\sin\theta$ is a spacetime volume factor, not a density; the proper volume element of the $t={\rm const}$ slices is $\alpha^{-1}\sqrt{-g}\,d^3x$, with $\alpha$ the lapse of the zero-angular-momentum observers~\cite{Bardeen:1972fi}.

\par
At the Kerr event horizon, the null generator is~\cite{Bardeen:1973gs,Wald:1984rg}
\begin{equation}
\xi=\partial_{t}+\Omega_{H}\,\partial_{\varphi} .
\label{eq-3.12}
\end{equation}
Because the radial derivative of $\xi\!\cdot\!\xi$ is negative at the outer horizon, $\xi$ is a future-directed ($\xi^{t}=1>0$) timelike vector just outside it. On a horizon cross section, the induced area element is $\sqrt\sigma=(r_{+}^{2}+a^{2})\sin\theta$~\cite{Bardeen:1973gs}. The total inward flux of the Dirac current across the future horizon $\mathcal H$ is defined as
\begin{equation}
\mathcal F=-\int_{\mathcal H}j^{\mu}\xi_{\mu}\,\sqrt\sigma\;dv\,d\theta\,d\varphi .
\label{eq-3.14}
\end{equation}
In the regular frame the current decomposes as $j=(|P_1'|^2+|Q_1'|^2)\,l'+(|P_0'|^2+|Q_0'|^2)\,n'+\dots$, where the dots denote terms along $m'$ and $\bar m'$ and the subscripts $0$ and $1$ label boost weights $+1/2$ and $-1/2$. For a field with a finite limit in this regular frame, $l'|_{\mathcal H}=\xi$ is null and orthogonal to $m'$, and the pointwise flux density across $\mathcal H^+$ is
\begin{equation}
-j_{\mu}\xi^{\mu}\big|_{\mathcal H}=|P_0'|^2+|Q_0'|^2\ \propto\ |Y_+|^2\big(S_{+1/2}^2+S_{-1/2}^2\big)\ \ge\ 0 .
\label{eq-3.15}
\end{equation}
Equivalently, with $\xi^{a}=\ell_\xi(1,\hat n)$ in a local orthonormal frame ($\ell_\xi>0$, $|\hat n|=1$), $-j^{\mu}\xi_{\mu}=\ell_\xi(j^{0}-\hat n\cdot\vec\jmath\,)\ge0$ by the causal bound. The flux density vanishes at a point of $\mathcal H$ if and only if the boost-weight $+1/2$ components vanish there, i.e., if the current is null and tangent to the generator, $j\propto\xi$. Such currents are possible with $\psi|_{\mathcal H}\neq0$, so positivity alone does not imply $\psi|_{\mathcal H}=0$; the field equations are needed.

\par
For separable fields, Eqs.~\eqref{eq-2.27}--\eqref{eq-2.28} with the source terms $\mathcal J_{1,2}$ imply the exact identity
\begin{eqnarray}
\frac{d\mathfrak F}{dr}&=&2\Delta^{1/2}\,{\rm Re}\big(\bar w_1\mathcal J_1-\bar w_2\mathcal J_2\big),\nonumber\\
\mathfrak F(r)&\equiv&\Delta^{1/2}\big(|w_1|^2-|w_2|^2\big)\nonumber\\
&=&|\Delta^{1/2}R_{+1/2}|^2-|R_{-1/2}|^2 .
\label{eq-flux}
\end{eqnarray}
This is the sphere-integrated continuity equation~\eqref{eq-cont}: up to a positive constant, $\mathfrak F(r)$ is the net flux of Dirac charge into the sphere $r={\rm const}$, and the right-hand side is the creation rate integrated over that sphere. Without source, $\mathfrak F$ is constant; it is the Wronskian of Chandrasekhar's radial system~\cite{Chandrasekhar:1984siy}. We have checked Eq.~\eqref{eq-flux} numerically (Appendix~\ref{app:repro}).

\par
Near $\mathcal H^+$ the homogeneous system has the two local solutions~\eqref{eq-2.31}. The ingoing one, $w_1\simeq c_{\rm in}\varepsilon^{-1/4-ik}$, $w_2=O(\varepsilon^{1/4-ik})$, has $Y_\pm$ analytic in $\varepsilon$. The outgoing one, $w_2\simeq c_{\rm out}\varepsilon^{-1/4+ik}$, $w_1=O(\varepsilon^{1/4+ik})$, has $Y_-\propto c_{\rm out}\,\varepsilon^{-1/2+2ik}$: its boost-weight $-1/2$ component diverges, the density seen by any freely falling observer diverges as $\varepsilon^{-1}$, and the charge on a regular slice through $\mathcal H^+$ diverges logarithmically. This holds for every real $k$, including $k=0$ (Sec.~\ref{sec:4.2}). Regularity on $\mathcal H^+$ therefore requires $c_{\rm out}=0$, and then $\mathfrak F(r_+)\propto|c_{\rm in}|^2>0$. For a bound state $\mathfrak F\to0$ at infinity; since $\mathfrak F$ is constant, $c_{\rm in}=0$, and the regular solution, which is fixed by $c_{\rm in}$ through the recursion~\eqref{eq-2.34}, vanishes identically. In particular, a current that is null and tangent to the generators requires $Y_+|_{\mathcal H}\propto c_{\rm in}=0$ and is therefore excluded; a non-vanishing boost-weight $-1/2$ component on $\mathcal H$ can only come from the singular outgoing branch. Hence $\psi\equiv0$ for regular, stationary, normalizable solutions of the homogeneous equation at every frequency, in agreement with Refs.~\cite{Finster:1998ak,Finster:1999ry}. Scattering states are not affected, since incoming waves from infinity balance the horizon flux~\cite{Brito:2015rjv}.

\par
All configurations computed here, and those of Ref.~\cite{Dzhunushaliev:2026rrt}, lie in the standing-wave subspace $w_2=\bar w_1$, $\mathcal J_2=\bar{\mathcal J}_1$ (Sec.~\ref{sec:4.2}), where $|w_1|=|w_2|$. Hence $\mathfrak F\equiv0$ and both sides of Eq.~\eqref{eq-flux} vanish for every $r>r_+$: the exterior limit of the net radial flux is zero, as is the creation rate integrated over each sphere (to round-off for our solutions; Appendix~\ref{app:repro}). The source therefore does not feed a net absorption by the horizon; its role is local. Without it, standing waves that decay at infinity exist for every $\omega$, but they contain the singular outgoing branch ($w_1\simeq c\,\varepsilon^{-1/4-ik}$, $w_2=\bar w_1$), and their static-frame amplitude and charge on the $t={\rm const}$ slices diverge. The singular source coefficients~\eqref{eq-4.8} replace this behavior by $w\to w^{(0)}$ (Case A) or $w\propto\varepsilon^{1/2}$ (Case B), which makes both finite and the spectrum discrete. 
\par
In the horizon-regular frame, Eq.~\eqref{eq-Y} gives $Y_+\propto\varepsilon^{1/4+ik}$ and $Y_-\propto\varepsilon^{-1/4+ik}$ for Case A. The density measured by freely falling observers diverges as $n_{\rm ff}\propto\varepsilon^{-1/2}$, which is locally integrable on a regular slice, but neither the spinor nor its current has a finite horizon value. Only the direction of the leading divergent current approaches the null generator, since its component along $l'$ grows as $\varepsilon^{-1/2}$. Case A therefore cannot be assigned a well-defined nonzero value of $\psi|_{\mathcal H}$. Its zero-flux statement refers to $\lim_{r\to r_+^+}\mathfrak F(r)=0$, not to a pointwise evaluation of Eq.~\eqref{eq-3.15} for a finite horizon field.

\par
For Case B, $Y_+\propto\varepsilon^{3/4+ik}$ and $Y_-\propto\varepsilon^{1/4+ik}$, so the exterior field extends continuously to zero and $n_{\rm ff}\propto\varepsilon^{1/2}$. However, $\partial_rY_-\propto(\tfrac14+ik)\varepsilon^{-3/4+ik}$ diverges for nonzero leading data. Thus Case B is not differentiable on $\mathcal H^+$ even at synchronization ($k=0$); for $k\neq0$ there is also logarithmic phase winding. Neither nontrivial sourced family is smooth on the future horizon. The ``horizon regularity'' used in prescribing these families means only finite static-frame radial amplitudes; finite integrated charge does not imply a smooth extension across $\mathcal H^+$.

\par
\begin{figure*}[t]
\centering
\includegraphics[width=14cm]{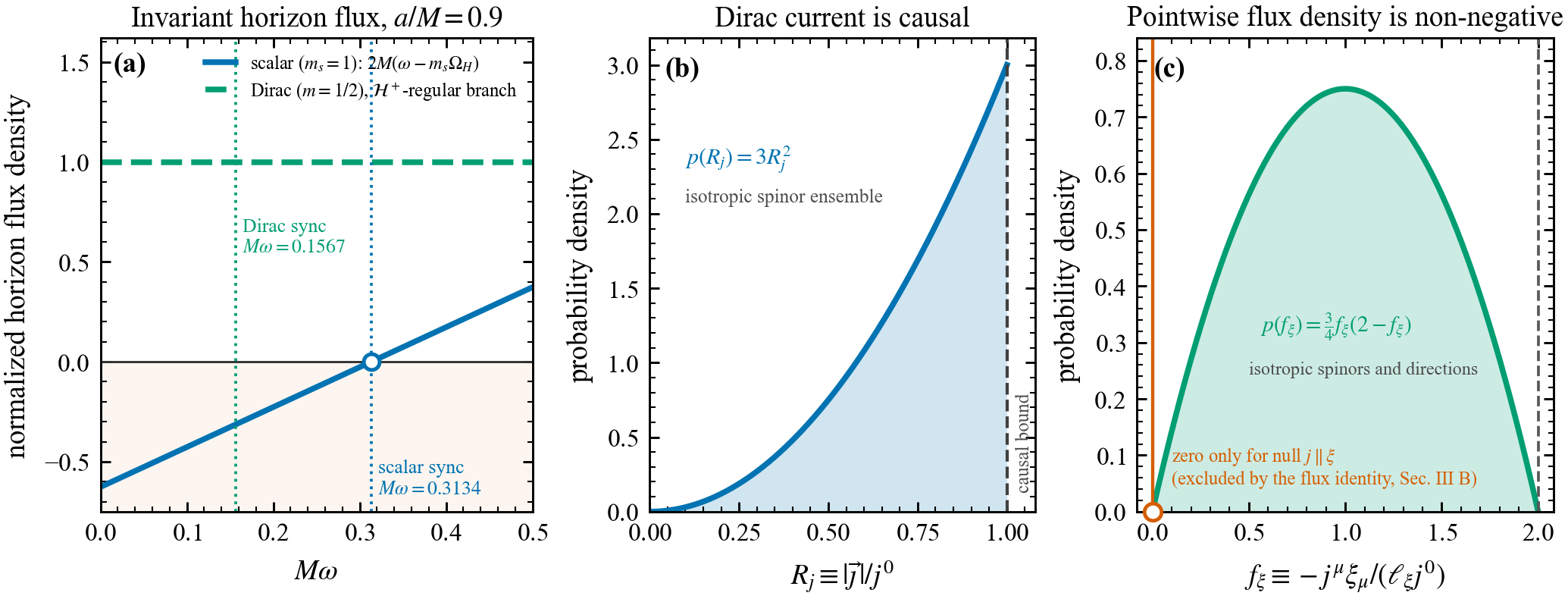}
\caption{Horizon flux of scalar and Dirac fields for $a/M=0.9$ ($M\Omega_H=0.3134$). (a) Invariant horizon flux densities, Eqs.~\eqref{eq-3.15} and~\eqref{eq-scalarflux}. Solid: scalar mode with $m_s=1$, $-N_\mu\xi^\mu/|\Phi|^2_{\mathcal H}=2(\omega-m_s\Omega_H)$ in units of $M^{-1}$, which changes sign at $M\omega=m_sM\Omega_H=0.3134$ (open circle). Dashed: the same ratio for the $\mathcal H^+$-regular Dirac branch ($m=1/2$), $-j_\mu\xi^\mu/(|P_0'|^2+|Q_0'|^2)=1$ [Eq.~\eqref{eq-3.15}]; the Dirac flux density has no kinematic prefactor and cannot change sign. Vertical dotted lines: scalar ($M\omega=0.3134$) and Dirac ($M\omega=mM\Omega_H=0.1567$) synchronization frequencies. (b) Distribution of $R_j=|\vec\jmath|/j^0$ for random spinors ($R_j\le1$). (c) Distribution of $f_{\xi}\equiv-j^{\mu}\xi_{\mu}/(\ell_\xi j^{0})$ for random spinors and null directions; it vanishes only for null currents along the generator, which are excluded for regular solutions by the flux identity~\eqref{eq-flux}.}
\label{fig:flux}
\end{figure*}

This structure distinguishes fermions from bosons. For a scalar field $\Phi=e^{-i\omega t+im\varphi}R(r)S(\theta)$, the $U(1)$ current $N_\mu=-i(\Phi^*\partial_\mu\Phi-\Phi\,\partial_\mu\Phi^*)$ has the Boyer--Lindquist charge density~\cite{Guo:2026pdi}
\begin{eqnarray}
N^{t} &=& \frac{2\omega}{\Sigma\Delta}
\Big[(r^{2}+a^{2})^{2}-\Delta a^{2}\sin^{2}\theta-\frac{2Mamr}{\omega}\Big]|\Phi|^{2} \nonumber\\
&\simeq & \frac{4Mr_{+}(r_{+}^{2}+a^{2})}{\Sigma_+\Delta}\big(\omega-m\Omega_{H}\big)|\Phi|^2 ,
\label{eq-3.17}
\end{eqnarray}
where the second line holds for $r\to r_+$, with $\Sigma_+\equiv\Sigma(r_+,\theta)$. This density diverges as $\Delta^{-1}$ because the Boyer--Lindquist slices degenerate at the horizon; the relevant quantity is the invariant flux density
\begin{equation}
-N_\mu\xi^\mu\big|_{\mathcal H}=2\big(\omega-m\Omega_H\big)|\Phi|^2_{\mathcal H} ,
\label{eq-scalarflux}
\end{equation}
which is finite for a field regular on $\mathcal H^+$. Because it is proportional to the kinematic detuning $(\omega-m\Omega_{H})$, the scalar flux density changes sign and vanishes at exact corotation. This kinematic zero enables bosonic superradiance~\cite{Zeldovich:1971ffh,Brito:2015rjv} and permits regular, sourceless scalar clouds at synchronization~\cite{Hod:2012px,Herdeiro:2014goa}. By contrast, the Dirac flux density~\eqref{eq-3.15} has no kinematic prefactor and is positive for the regular branch at all frequencies, so fermions cannot quench their horizon flux (Fig.~\ref{fig:flux}).

\par
In summary, exact corotation ($k=0$) removes the Boyer--Lindquist phase winding and produces no logarithmic branch, but it changes neither the selection of the ingoing solution by regularity on $\mathcal H^+$ nor the positivity of its flux. Synchronization therefore opens no existence window for source-free Dirac bound states, unlike for scalar clouds~\cite{Hod:2012px,Herdeiro:2014goa}; consistently, Dirac quasi-bound states of Kerr black holes decay at all frequencies~\cite{Dolan:2015eua}.

\section{Global Solutions and Physical Properties of Rotating Sourced Bound States}
\label{sec:4}
\par
\subsection{Angular Eigenvalues and Morphology of the Field Density}
\label{sec:4.1}
\par
Unlike scalar fields, which couple to rotation via the consolidated parameter $c^{2}=a^{2}(\omega^{2}-\mu^{2})$~\cite{Teukolsky:1973ha,Berti:2005gp,Brito:2015rjv}, the Chandrasekhar--Page angular system [Eqs.~\eqref{eq-2.18}--\eqref{eq-2.19}] depends on two independent couplings: $a\omega$ and $a\tilde m$. Assuming $m>0$, the non-rotating limit $a\to0$ simplifies the operator $Q$ to $m/\sin\theta$, yielding the exact spin-weighted spherical spectrum~\cite{Page:1976jj,Chandrasekhar:1984siy}:
\begin{equation}
\lambda=\pm\Big(j+\tfrac12\Big) ,
\qquad j=|m|,\,|m|+1,\,\dots
\label{eq-4.1}
\end{equation}

\par
For Dirac fields, the azimuthal number $m$ is half-integer. The fundamental mode corresponds to $j=\tfrac12$ ($|\lambda|=1$), with the negative branch $\lambda=-1$ selected by the spherical mapping (Sec.~\ref{sec:4.2}). Near the polar singularity ($\theta\to0$), indicial analysis of $Q\pm\tfrac12\cot\theta$ yields the regular boundary branches~\cite{Chandrasekhar:1984siy}:
\begin{eqnarray}
S_{-1/2} &=& \theta^{\,m-1/2}\big(1+\mathcal{O}(\theta^{2})\big) , \nonumber\\
S_{+1/2} &=& \frac{a\tilde m-\lambda}{2m+1}\,\theta^{\,m+1/2}\big(1+\mathcal{O}(\theta^{2})\big) .
\label{eq-4.2}
\end{eqnarray}
The leading coefficient of $S_{+1/2}$ demonstrates that mass coupling modifies the near-axis structure. Omitting this term from the start-up data of a shooting integration introduces a truncation error that grows with the start-up angle $\theta_0$ (about $5\times10^{-13}$ in $\lambda$ for $\theta_0=10^{-6}$ in our code).

\par
\begin{figure*}[htbp]
\centering
\includegraphics[width=13.5cm]{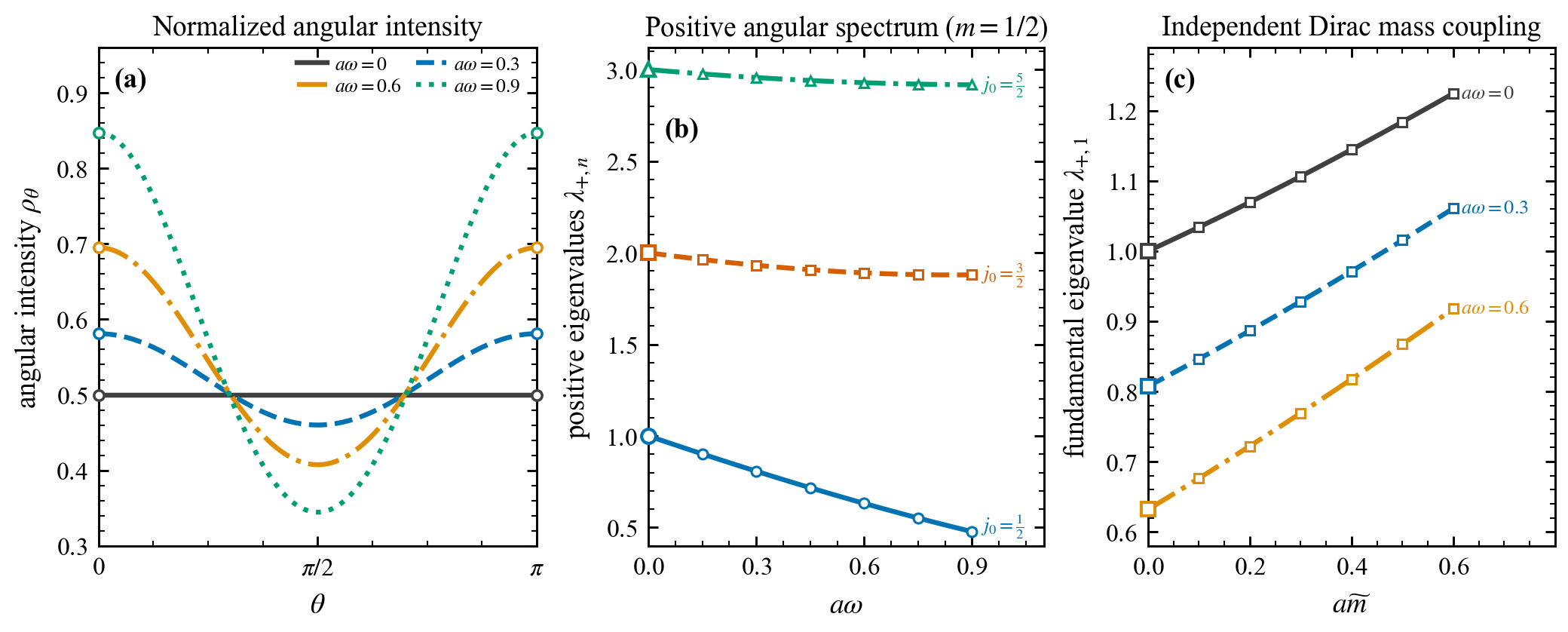}
\caption{The angular eigenvalue problem for $M=1$. (a) Normalized angular intensity $\rho_\theta=S_{+1/2}^2+S_{-1/2}^2$ of the fundamental mode ($m=\tfrac12$, $\lambda_+$ branch, $a\tilde m=0$) for several values of $a\omega$. Since $S_{-1/2}(0)\neq0$, the density remains finite on the rotation axis for $|m|=1/2$; for $|m|\ge3/2$ it vanishes there [Fig.~\ref{fig:shape}(c)]. (b) The three lowest positive eigenvalues $\lambda_{+,n}$ versus $a\omega$ ($a\tilde m=0$). Kinetic rotation lowers the eigenvalue from the exact $a=0$ values (large open symbols). (c) Fundamental eigenvalue $\lambda_{+,1}$ versus $a\tilde m$ for $a\omega=0$, $0.3$, and $0.6$. Mass coupling opposes the kinetic effect.}
\label{fig:angular}
\end{figure*}
\begin{figure*}[htbp]
\centering
\includegraphics[width=14cm]{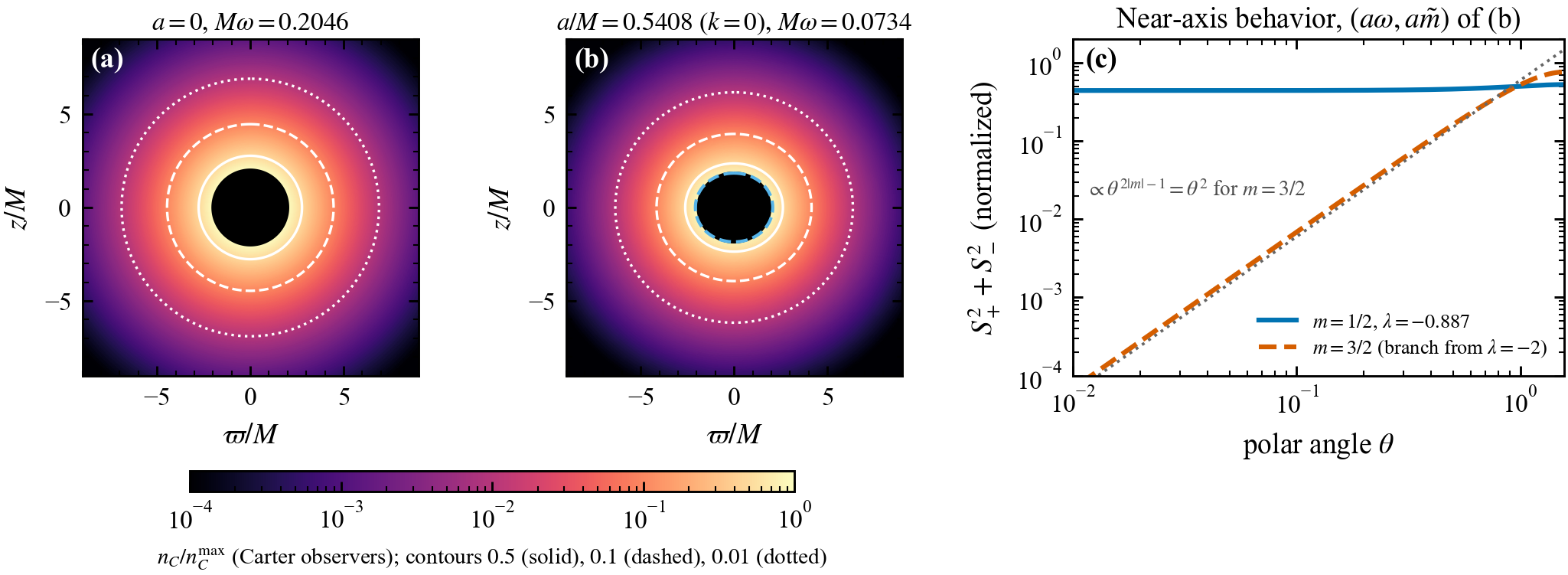}
\caption{Reconstructed density (Case A, $v_0/u_0=-0.2$, $m=1/2$, $M\tilde m=0.5$, self-consistent $\lambda$). (a) Carter-observer density $n_C$, Eq.~\eqref{eq-4.7}, for $a=0$ ($M\omega=0.2046$), on a logarithmic scale normalized to its maximum; white contours: $n_C/n_C^{\max}=0.5$ (solid), $0.1$ (dashed), $0.01$ (dotted). (b) The same at the synchronization point $a/M=0.5408$ ($M\omega=0.0734$); black: horizon; cyan dashed: stationary limit; coordinates $\varpi=\sqrt{r^2+a^2}\sin\theta$, $z=r\cos\theta$. (c) $S_{+1/2}^2+S_{-1/2}^2$ near the axis for the $m=1/2$ configuration of (b) and for an $m=3/2$ mode with the same $(a\omega,a\tilde m)$, which vanishes as $\theta^2$.}
\label{fig:shape}
\end{figure*}

Equation~\eqref{eq-4.2} dictates a near-axis scaling $\theta^{m\mp1/2}$. For the fundamental mode $m=\tfrac12$, the component $S_{-1/2}\sim\theta^{0}$ remains finite [Fig.~\ref{fig:angular}(a)], so the density does not vanish on the rotation axis. This holds only for $|m|=1/2$: for $|m|\ge3/2$, $S_{-1/2}\sim\theta^{|m|-1/2}$ and $S_{+1/2}\sim\theta^{|m|+1/2}$, and the density vanishes on the axis as $\theta^{2|m|-1}$ [Fig.~\ref{fig:shape}(c)]. A half-integer azimuthal number alone therefore does not exclude configurations that are hollow along the axis. The difference from scalar fields is narrower: co-rotating scalar modes need $m\ge1$ and vanish on the axis because of the centrifugal barrier $\sim m^{2}/\sin^{2}\theta$~\cite{Hod:2012px,Herdeiro:2014goa,Guo:2026pdi}, whereas $|m|=1/2$ Dirac modes can co-rotate and remain finite there. A finite density on the axis does not fix the global shape either; the energy distribution of self-gravitating spinning Dirac stars, for example, is toroidal~\cite{Herdeiro:2019mbz}. We therefore determine the shape of our configurations by full reconstruction below. Globally, the angular system possesses an exact parity symmetry~\cite{Page:1976jj} despite the apparent asymmetry of the $a\tilde m\cos\theta$ term:
\begin{equation}
(S_{+1/2},S_{-1/2})(\theta)\ \longrightarrow\ (S_{-1/2},S_{+1/2})(\pi-\theta) .
\label{eq-4.3}
\end{equation}
This symmetry enables equatorial matching in our shooting scheme. We independently verified the spectrum using a matrix spectral method based on spin-weighted spherical harmonics of spin weight $\pm1/2$, along the lines of Ref.~\cite{Dolan:2009kj}. Both techniques agree to $\mathcal{O}(10^{-13})$ (Appendix~\ref{app:repro}), which rules out a systematic solver bias.

\par
Perturbative analysis at small couplings quantifies the competing effects of rotation and mass:
\begin{eqnarray}
\lambda_{1} &=& 1-\tfrac23\,a\omega+\tfrac{2}{27}\,(a\omega)^{2}+\mathcal{O}\big((a\omega)^3\big) , \nonumber\\
\lambda_{1} &=& 1+\tfrac13\,a\tilde m+\tfrac{2}{27}\,(a\tilde m)^{2}+\mathcal{O}\big((a\tilde m)^3\big) .
\label{eq-4.4}
\end{eqnarray}
Kinetic rotation lowers the eigenvalue, while mass coupling restores it. This competition persists at finite couplings (Fig.~\ref{fig:angular}), and our numerical mixed derivative recovers the exact cross-coupling coefficient:
\begin{equation}
\frac{\partial^{2}\lambda_{1}}{\partial(a\omega)\,\partial(a\tilde m)}=\frac{4}{27} .
\label{eq-4.5}
\end{equation}
Equations~\eqref{eq-4.4}--\eqref{eq-4.5} refer to the positive branch $\lambda_+$ that starts at $\lambda=1$ (a double series in $a\omega$ and $a\tilde m$ is given in Ref.~\cite{Dolan:2009kj}). Since Eqs.~\eqref{eq-2.18}--\eqref{eq-2.19} are invariant under $(\lambda,\tilde m,S_{-1/2})\to(-\lambda,-\tilde m,-S_{-1/2})$, the negative branch used in Sec.~\ref{sec:4.2} obeys $\lambda_-(a\omega,a\tilde m)=-\lambda_+(a\omega,-a\tilde m)=-1+\tfrac23a\omega+\tfrac13a\tilde m+\dots$, so rotation and mass both reduce $|\lambda_-|$.

\par
\begin{figure*}[htbp]
\centering
\includegraphics[width=12cm]{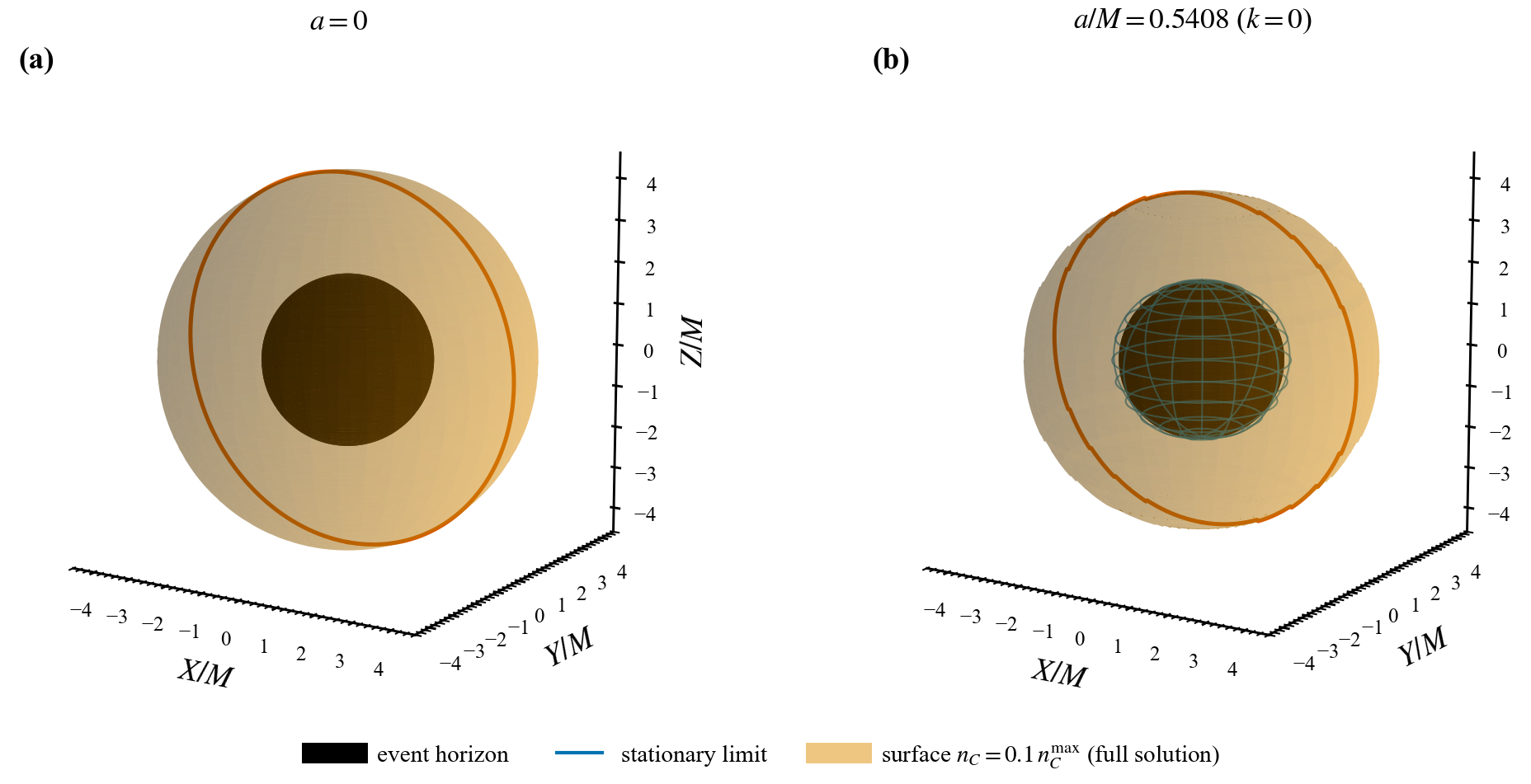}
\caption{Isodensity surfaces $n_C=0.1\,n_C^{\max}$ (orange) of the configurations of Fig.~\ref{fig:shape}: (a) $a=0$; (b) $a/M=0.5408$ ($k=0$). Black: event horizon; blue mesh: stationary limit. Coordinates: $X=\sqrt{r^2+a^2}\sin\theta\cos\varphi$, $Y=\sqrt{r^2+a^2}\sin\theta\sin\varphi$, $Z=r\cos\theta$.}
\label{fig:3d}
\end{figure*}

In the highly prolate regime ($a\omega \gg 1$), the effective potential wells near the rotational poles decouple, dropping the fundamental eigenvalue exponentially. Using multi-precision arithmetic, our spectral solver numerically identifies the asymptotic scaling $\lambda_{1}\simeq 4\,a\omega\,e^{-2a\omega}$, tracking $\lambda_{1}$ down to $10^{-33}$ at $a\omega=40$. Standard shooting methods fail in this limit. For this mode ($m=j$), whose eigenvalue tends to zero as $a\omega\to\infty$, Ref.~\cite{Dolan:2009kj} found no satisfactory analytic large-$a\omega$ asymptote. Such eigenvalues enter high-frequency Dirac perturbations of rotating black holes, whose quasinormal spectra are computed, e.g., in Ref.~\cite{Jing:2005pk}. Regarding the shape of the configurations: the trace over the non-rotating spinor components yields a purely spherical density profile~\cite{Dzhunushaliev:2026rrt}:
\begin{equation}
\sum_{\beta}|\chi_{\beta}|^{2}
=\tfrac14\big(2|\tilde u|^{2}+2|\tilde v|^{2}\big)
=\tfrac12\big(|\tilde u|^{2}+|\tilde v|^{2}\big) .
\label{eq-4.6}
\end{equation}
In Kerr the density has to be reconstructed from the full solution. Boosting the Kinnersley components to the Carter frame~\eqref{eq-carter} gives component moduli $\Sigma^{-1/4}|w_{1,2}||S_{\pm1/2}|$ up to a common constant, so the number density measured by the Carter observers is
\begin{eqnarray}
n_C&=&-j_\mu e_0^\mu\ \propto\ \frac{|w_1|^2+|w_2|^2}{2}\,\frac{S_{+1/2}^2+S_{-1/2}^2}{\sqrt\Sigma}\nonumber\\
&=&|w(r)|^2\,\frac{S_{+1/2}^2(\theta)+S_{-1/2}^2(\theta)}{\sqrt{\Sigma(r,\theta)}} ,
\label{eq-4.7}
\end{eqnarray}
where the last form holds in the subspace $w_2=\bar w_1$ used for all our solutions; for $a=0$ it reduces to Eq.~\eqref{eq-4.6}, since $S_{+1/2}^2+S_{-1/2}^2$ is constant for $j=|m|=1/2$. Figures~\ref{fig:shape} and~\ref{fig:3d} show $n_C$ for the benchmark ($a=0$) and for the self-consistent Kerr configuration of the $v_0/u_0=-0.2$ branch at its synchronization point, $a/M=0.5408$ ($M\omega=0.07343$, $\lambda=-0.8875$), computed from the radial solution of Sec.~\ref{sec:4.2} and the spectral angular eigenfunctions. The Kerr configuration is only mildly oblate: the ratio of the equatorial to the polar extent of the $n_C/n_C^{\max}=0.5$, $0.1$, and $0.01$ contours is $1.10$, $1.05$, and $1.03$, and near the horizon, where $n_C$ peaks, the equatorial density exceeds the polar one by $24\%$. The density is finite on the axis, while an $|m|=3/2$ mode with the same couplings vanishes there as $\theta^2$ [Fig.~\ref{fig:shape}(c)]. Since the density was reconstructed only for the baseline branches of Sec.~\ref{sec:4.2}, which exist only for $a\lesssim0.66M$, stronger flattening at larger spins was not examined, and no general statement about the shape of fermionic configurations follows from these solutions.

\subsection{The Sourced Radial Problem and the Kinematic Synchronization Limit}
\label{sec:4.2}
\par
Matching the prescribed static-frame asymptotics fixes the singular coefficient of the external source near the event horizon. Substituting $w_{1,2}=w^{(0)}_{1,2}\varepsilon^{p}$ into the near-horizon equations (Eqs.~\eqref{eq-2.29}--\eqref{eq-2.30}) with sources $\mathcal J_{1,2}$ yields at $\mathcal{O}(\varepsilon^{p-1})$:
\begin{equation}
\big[\mathcal J_{1,2}\big]_{\varepsilon^{\,p-1}}
=\Big(\tfrac14+p \pm ik\Big)\,w^{(0)}_{1,2} ,
\label{eq-4.8}
\end{equation}
where the upper and lower signs correspond to $\mathcal J_{1}$ and $\mathcal J_{2}$. Here, $p=0$ (finite static-frame amplitude, Case A) represents a first-order pole, whereas $p=\tfrac12$ (vanishing static-frame amplitude, Case B) defines a branch-point coefficient~\cite{Dzhunushaliev:2026rrt}. The quantities $w^{(0)}_{1,2}$ are leading coefficients of an exterior expansion, not values of the spinor in a horizon-regular frame. As $a\to0$ ($k\to\Omega$), mapping $w_{1}\leftrightarrow r^{1/2}(v+iu)$ recovers the spherical constraints of Eqs.~\eqref{eq-3.4} and~\eqref{eq-3.5} (as we have checked symbolically).

\par
\begin{figure*}[htbp]
\centering
\includegraphics[width=12cm]{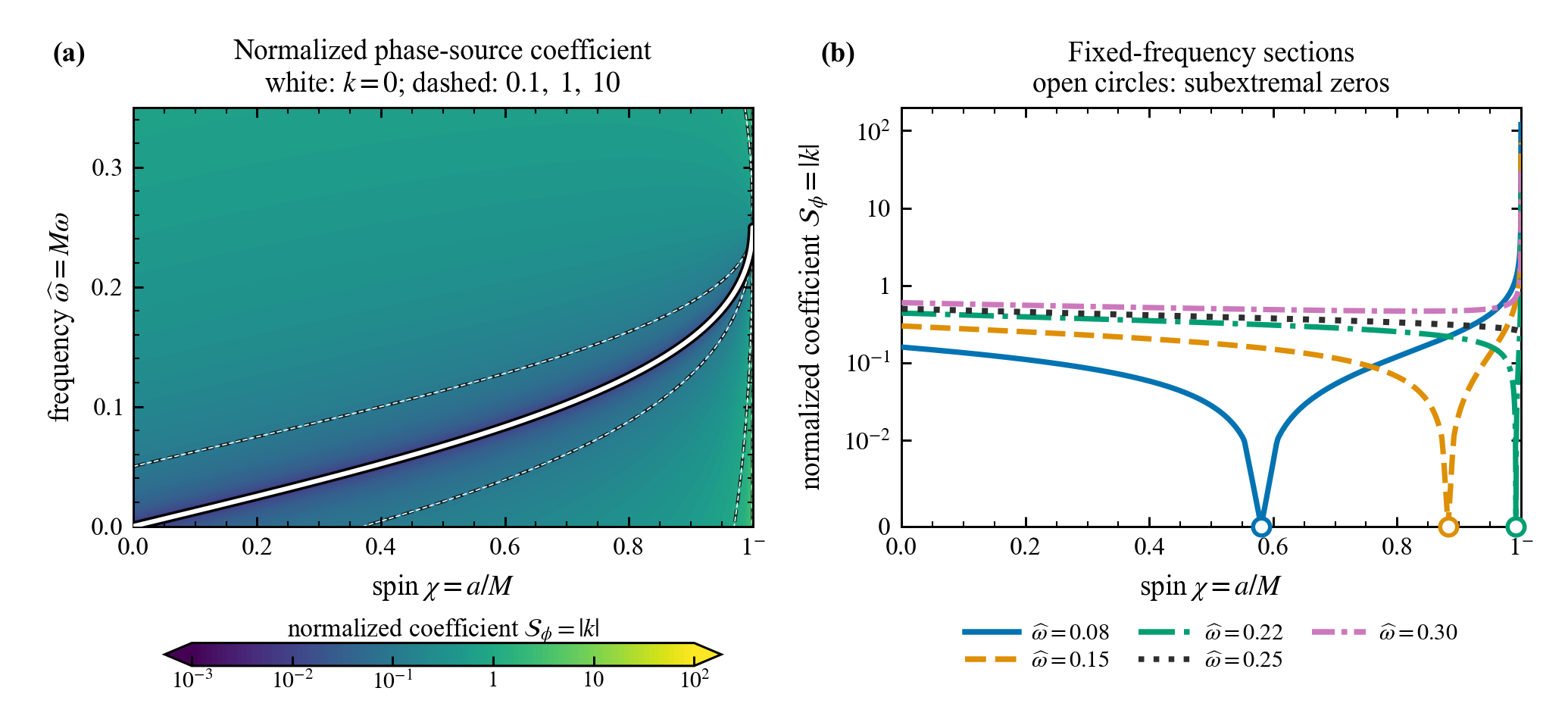}
\caption{The phase-winding part $|k|$ of the singular source coefficient [Eqs.~\eqref{eq-4.8}--\eqref{eq-4.9}], normalized by the leading static-frame coefficient $|w^{(0)}|$, for $m=\tfrac12$. (a) $|k|$ over the $(a/M,M\omega)$ plane; it vanishes on the synchronization locus $\omega=m\Omega_{H}$ (solid white line), and the dashed lines mark $|k|=0.1$, $1$, and $10$. (b) Sections at fixed $M\omega$; open circles mark the zeros on the synchronization locus. The constant weight part $\tfrac14+p$ is not shown.
\label{fig:source}}
\end{figure*}

The two terms in Eq.~\eqref{eq-4.8} have different origins. The term $(\tfrac14+p)w^{(0)}$ compensates the frame weight $\Delta'/(4\Delta)$, i.e., the spin-connection term of the Carter frame, together with the prescribed power $\varepsilon^{p}$; the term $\pm ik\,w^{(0)}$ compensates the Boyer--Lindquist phase winding. The rescaling $\hat w_{1,2}=\Delta^{1/4}w_{1,2}$ removes the weight. It gives the Kinnersley variables $\hat w_1=\Delta^{1/2}R_{+1/2}$ and $\hat w_2=R_{-1/2}$, reduces to $\hat v=f^{1/4}v$, $\hat u=f^{1/4}u$ for $a=0$ (Sec.~\ref{sec:3.1}), and turns Eq.~\eqref{eq-2.28} into $\hat w_1'+(iK/\Delta)\hat w_1=(\lambda-i\tilde mr)\Delta^{-1/2}\hat w_2+\Delta^{1/4}\mathcal J_1$ (similarly for $\hat w_2$), so that for $\hat w_{1,2}=\hat w^{(0)}_{1,2}\varepsilon^{\hat p}$
\begin{equation}
\big[\Delta^{1/4}\mathcal J_{1,2}\big]_{\varepsilon^{\hat p-1}}=\big(\hat p\pm ik\big)\,\hat w^{(0)}_{1,2},\qquad \hat p=p+\tfrac14 .
\label{eq-4.9}
\end{equation}
Only for a finite Kinnersley amplitude ($\hat p=0$, i.e., $w\propto\varepsilon^{-1/4}$ as for the homogeneous solutions) is the singular coefficient purely imaginary and zero at synchronization. In Cases A and B ($\hat p=1/4$ and $3/4$) only the phase-winding part vanishes at $\omega=m\Omega_H$, and $|[\mathcal J_1]_{\varepsilon^{p-1}}|/|w^{(0)}|=|\tfrac14+p+ik|$ reaches its minimum $\tfrac14+p$, not zero. The phenomenological source amplitude is thus geometrically locked; its phase-winding part is proportional to the invariant $k$ ($\propto|\omega-m\Omega_{H}|/\kappa_{+}$)~\cite{Brito:2015rjv}, as shown in Fig.~\ref{fig:source}.

\par
For real parameters $(\omega, \lambda, \tilde m)$, the system admits the invariant subspace $w_{2}=\overline{w_{1}}$~\cite{Chandrasekhar:1984siy} (with $\mathcal J_2=\overline{\mathcal J_1}$), reducing the problem to a single complex first-order equation for $w\equiv w_{1}$:
\begin{equation}
\frac{dw}{dr}+\Big[\frac{\Delta'}{4\Delta}+\frac{iK}{\Delta}\Big]w
=\big(\lambda-i\tilde mr\big)\Delta^{-1/2}\,\overline{w}+J,\qquad J\equiv\mathcal J_1 .
\label{eq-4.10}
\end{equation}
Mapping to the spherically symmetric ansatz fixes the parameters:
\begin{equation}
w=r^{1/2}\,(v+iu) ,\qquad \lambda=-1 ,\qquad
\tilde m=\tilde m_{\rm DF} ,
\label{eq-4.11}
\end{equation}
which selects the fundamental negative angular branch ($|\lambda|=1$); here $\tilde m_{\rm DF}\equiv m/r_H$ is the spinor mass in the notation of Sec.~\ref{sec:3.1}, and with this map Eq.~\eqref{eq-4.10} reproduces Eqs.~\eqref{eq-3.1}--\eqref{eq-3.2} exactly, with $J=r^{1/2}f^{-1/2}(j_1+ij_2)$. The required covariant source profile translates to:
\begin{eqnarray}
&&J(r)=\frac{r}{r_{+}}\,e^{-\Sigma(r-r_{+})}
\Big[\frac{\rho}{r-r_{+}}+\frac{\eta}{\sqrt{r-r_{+}}}\Big] , \nonumber \\
&&\Sigma=\sqrt{\tilde m^{2}-\omega^{2}} .
\label{eq-4.12}
\end{eqnarray}
For Case A, define $w^{(0)}\equiv\lim_{\varepsilon\to0^+}w(r_++\varepsilon)$, the finite limit of the static-frame radial amplitude. The coefficients are $\rho=(\tfrac14+ik)\,w^{(0)}$, fixed by Eq.~\eqref{eq-4.8}, and $\eta=-(\lambda-i\tilde mr_+)(r_+-r_-)^{-1/2}\,\overline{w^{(0)}}$, fixed by the absence of an $O(\varepsilon^{1/2})$ term; for $a=0$ they reproduce the coefficients of Eq.~\eqref{eq-3.4} (Appendix~\ref{app:repro}). In Case B the bracket is replaced by $\rho'/\sqrt{r-r_+}+\eta'$, with $\rho'=(\tfrac34+ik)\,w^{(1)}$ and $\eta'=-(\lambda-i\tilde mr_+)(r_+-r_-)^{-1/2}\,\overline{w^{(1)}}$, where $w\simeq w^{(1)}\varepsilon^{1/2}$. Thus the source coefficients are fixed by the prescribed static-frame expansion data, without assigning a finite horizon value to the Case A spinor. Our numerical integration reproduces the spherical benchmark $\Omega_{\rm A}=0.409128952$ (Sec.~\ref{sec:3.1}) as $a\to0$.

\par
\begin{figure*}[htbp]
\centering
\includegraphics[width=13.5cm]{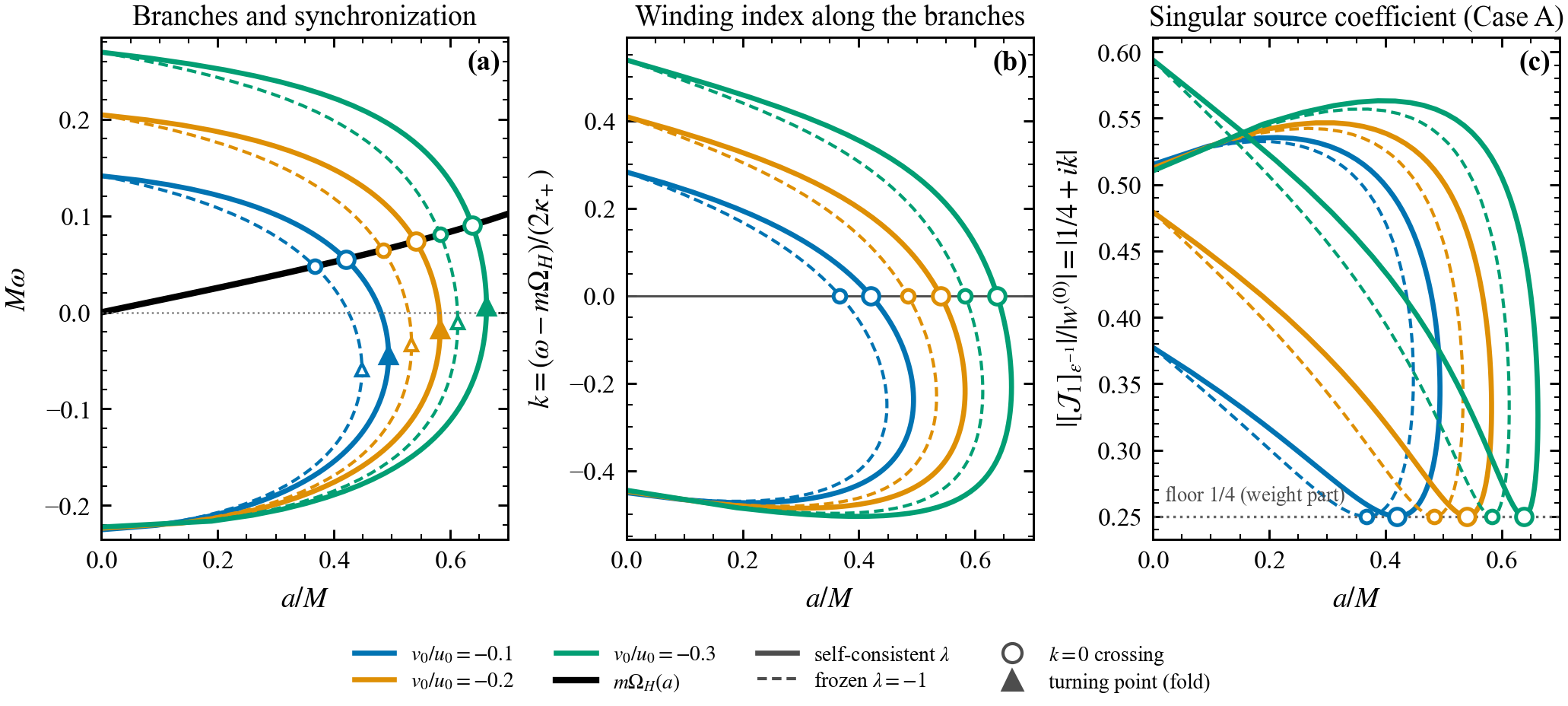}
\caption{Branch structure for $m=\tfrac12$ and $M\tilde m=0.5$ (Case A). (a) Frequency of the branches $v_0/u_0=-0.1$, $-0.2$, and $-0.3$ (colors) for self-consistent (solid) and frozen (dashed) $\lambda$, with the synchronization line $m\Omega_{H}$ (black). Each arc starts at the benchmark-connected root, crosses the synchronization line transversally (open circles), and turns back at a fold close to $\omega=0$ (triangles; filled: self-consistent, open: frozen) into the negative-frequency branch. (b) Winding index $k$ along the arcs. (c) Modulus of the singular source coefficient, $|[\mathcal J_1]_{\varepsilon^{-1}}|/|w^{(0)}|=|\tfrac14+ik|$, which reaches its floor $1/4$ at synchronization.
\label{fig:scan}}
\end{figure*}

For the baseline source~\eqref{eq-4.12}, sweeping $a$ for $m=\tfrac12$ and $M\tilde m=0.5$, we search for roots over the whole interval $-\tilde m<\omega<\tilde m$ and, near turning points, continue in $\omega$ and solve for $a(\omega)$ (Appendix~\ref{app:repro}). The results are shown in Fig.~\ref{fig:scan} and Table~\ref{tab:scan}. The frequency of the branch connected to the benchmark decreases monotonically and crosses the synchronization line $m\Omega_{H}(a)$ transversally at $a_{\rm sync}$, where $k$ changes sign; nothing attracts the frequency to this line. At the crossing the phase-winding part of the singular coefficient vanishes and the coefficient reaches its floor $1/4$ [Fig.~\ref{fig:scan}(c)]. The branch does not end at $\omega\to0$: it reaches a turning point at $a_{\rm fold}$, close to $\omega=0$ ($|M\omega|<0.06$), where it merges with a second solution that starts at $a=0$ at a negative frequency ($M\omega=-0.2252$, $-0.2236$, and $-0.2224$ for $v_0/u_0=-0.1$, $-0.2$, and $-0.3$; cf.\ Sec.~\ref{sec:3.1}). For $a>a_{\rm fold}$ no further root was found in the scanned frequency interval for this source prescription and these asymptotic data. The smooth continuation in $\omega$ through the turning point supports a saddle-node interpretation for these two branches.

\par
Freezing $\lambda=-1$ is adequate for small couplings; in general $\lambda$ must be evaluated self-consistently, $\lambda=\lambda_-(a\omega,a\tilde m)$, on the branch connected to $\lambda=-1$ and with the same $\tilde m$ as in the radial equation. Rather than iterating $\omega\to(a\omega,a\tilde m)\to\lambda\to\omega$~\cite{Guo:2026pdi}, we find the roots of the combined residual, recomputing $\lambda$ spectrally at each trial frequency, which avoids the fragility of the iteration near $\omega\to0$. Since $\lambda_-=-1+\tfrac23a\omega+\tfrac13a\tilde m+\dots$ (Sec.~\ref{sec:4.1}), the self-consistent coupling reduces $|\lambda|$ and shifts $a_{\rm sync}$ and $a_{\rm fold}$ to larger spins (Table~\ref{tab:scan}); the transversal crossing and the fold persist.
\begin{table*}[t]
\centering
\caption{Branches of the $\omega(a)$ scan for Case A ($M=1$, $m=\tfrac12$, $\tilde m=0.5$). $\Omega^{\pm}(a{=}0)=2M\omega$ ($r_H$ units) are the benchmark-connected root and the negative-frequency root with the same horizon data; $a_{\rm sync}$ is the crossing of $\omega=m\Omega_{H}$, and $a_{\rm fold}$ the turning point where the two branches merge. Frozen: $\lambda=-1$; self-cons.: $\lambda=\lambda_-(a\omega,a\tilde m)$; $\lambda_{\rm sync}$ is the self-consistent eigenvalue at $a_{\rm sync}$. The entries are numerical estimates for the baseline source profile; representative frequency sensitivity tests and their scope are given in Appendix~\ref{app:repro}.
\label{tab:scan}}
\begin{tabular}{cccccccc}
\hline\hline
 & & & \multicolumn{2}{c}{$a_{\rm sync}$} & \multicolumn{2}{c}{$a_{\rm fold}$} & \\
$v_{0}/u_{0}$ & $\Omega^{+}(a{=}0)$ & $\Omega^{-}(a{=}0)$ & frozen & self-cons. & frozen & self-cons. & $\lambda_{\rm sync}$ \\
\hline
$-0.1$ & $0.282631$ & $-0.450428$ & $0.3673$ & $0.4205$ & $0.4484$ & $0.4941$ & $-0.9171$ \\
$-0.2$ & $0.409129$ & $-0.447269$ & $0.4844$ & $0.5408$ & $0.5338$ & $0.5829$ & $-0.8875$ \\
$-0.3$ & $0.538767$ & $-0.444875$ & $0.5833$ & $0.6383$ & $0.6133$ & $0.6627$ & $-0.8605$ \\
\hline\hline
\end{tabular}
\end{table*}

\par
\begin{figure*}[htbp]
\centering
\includegraphics[width=13.5cm]{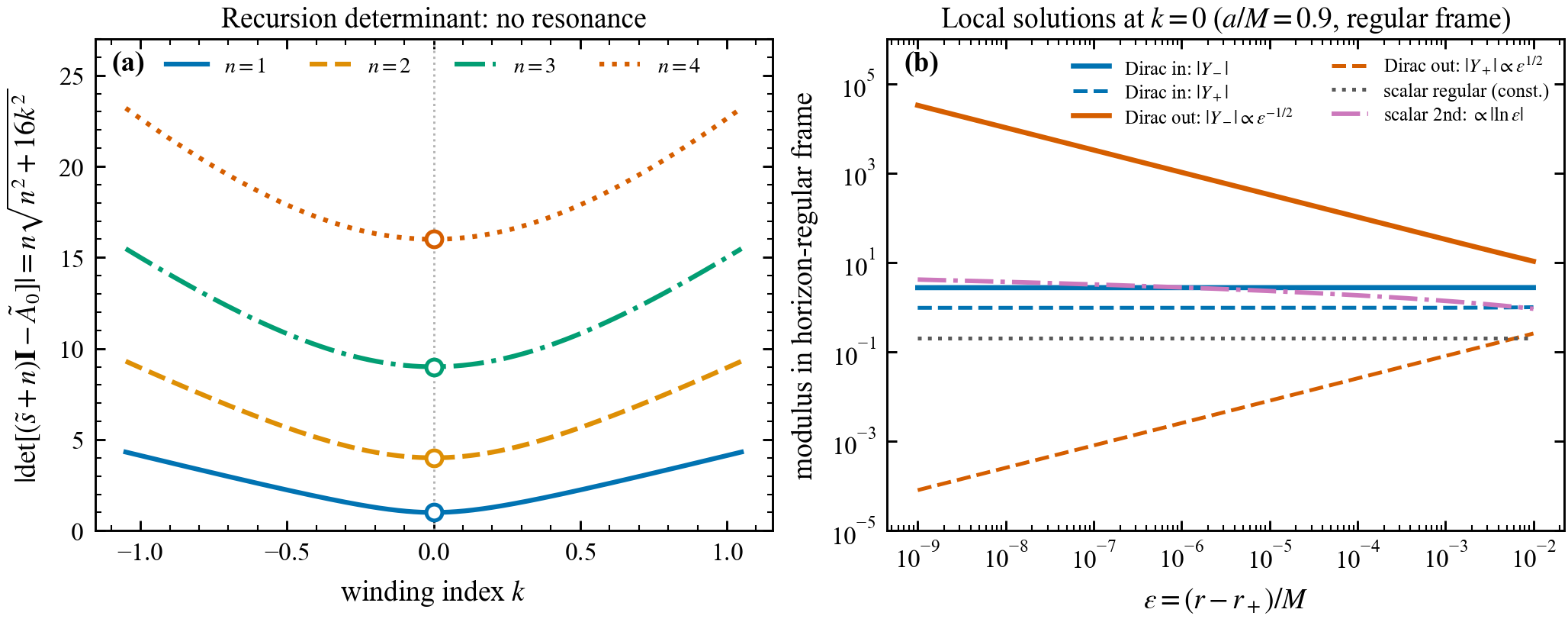}
\caption{The Frobenius structure at $k=0$. (a) $|\det[(\tilde s+n)\mathbf{I}_2-\tilde A_{0}]|=n\sqrt{n^2+16k^2}$, which equals $n^{2}\neq0$ at synchronization (open circles), so no logarithms appear. (b) Moduli of the two local solutions at $k=0$ in the horizon-regular frame, Eq.~\eqref{eq-Y}, from the Frobenius series ($a/M=0.9$, $m=\tfrac12$, $M\tilde m=0.5$, $\lambda=-1$). The ingoing Dirac solution is regular ($|Y_\pm|\to$ const), whereas the outgoing one has $|Y_-|\propto\varepsilon^{-1/2}$ and $|Y_+|\propto\varepsilon^{1/2}$; the two scalar solutions behave as a constant and as $|\ln\varepsilon|$ (arbitrary normalization). In both cases regularity on $\mathcal H^+$ selects one solution.
\label{fig:degeneracy}}
\end{figure*}

At exact synchronization ($k=0$), the leading Frobenius matrix degenerates to $\tilde A_{0}=-\tfrac12\mathbf{I}_2$. Being proportional to the identity (non-defective) with a non-zero recursion determinant $n^{2}$, no resonances or logarithmic branches appear. Only the ingoing solution, $a_0\propto(1,0)^{\mathsf T}$, is regular on $\mathcal H^+$ (Fig.~\ref{fig:degeneracy}), and by the flux identity of Sec.~\ref{sec:3.2} it cannot describe a bound state unless $a_0=0$, in which case all $a_n$ vanish. Sourceless, stationary, bound, corotating Dirac configurations regular on $\mathcal H^+$ therefore do not exist in the separated framework, consistent with Refs.~\cite{Finster:1998ak,Finster:1999ry}.

\par
This highlights a fermion--boson dichotomy. For scalars, the Teukolsky equation produces a defective Jordan block at corotation, generating a logarithmically divergent second solution, which regularity removes; the remaining solution carries the flux $2(\omega-m\Omega_H)|\Phi|^2=0$, so the decay condition at infinity can be met for discrete parameters, which gives the synchronized bosonic clouds~\cite{Hod:2012px,Herdeiro:2014goa}. For the Dirac field regularity also selects a single solution, but this solution carries a positive flux. What matters is therefore the flux of the regular solution, not the presence or absence of a logarithm (Fig.~\ref{fig:degeneracy}).

\par
\begin{figure*}[htbp]
\centering
\includegraphics[width=12.5cm]{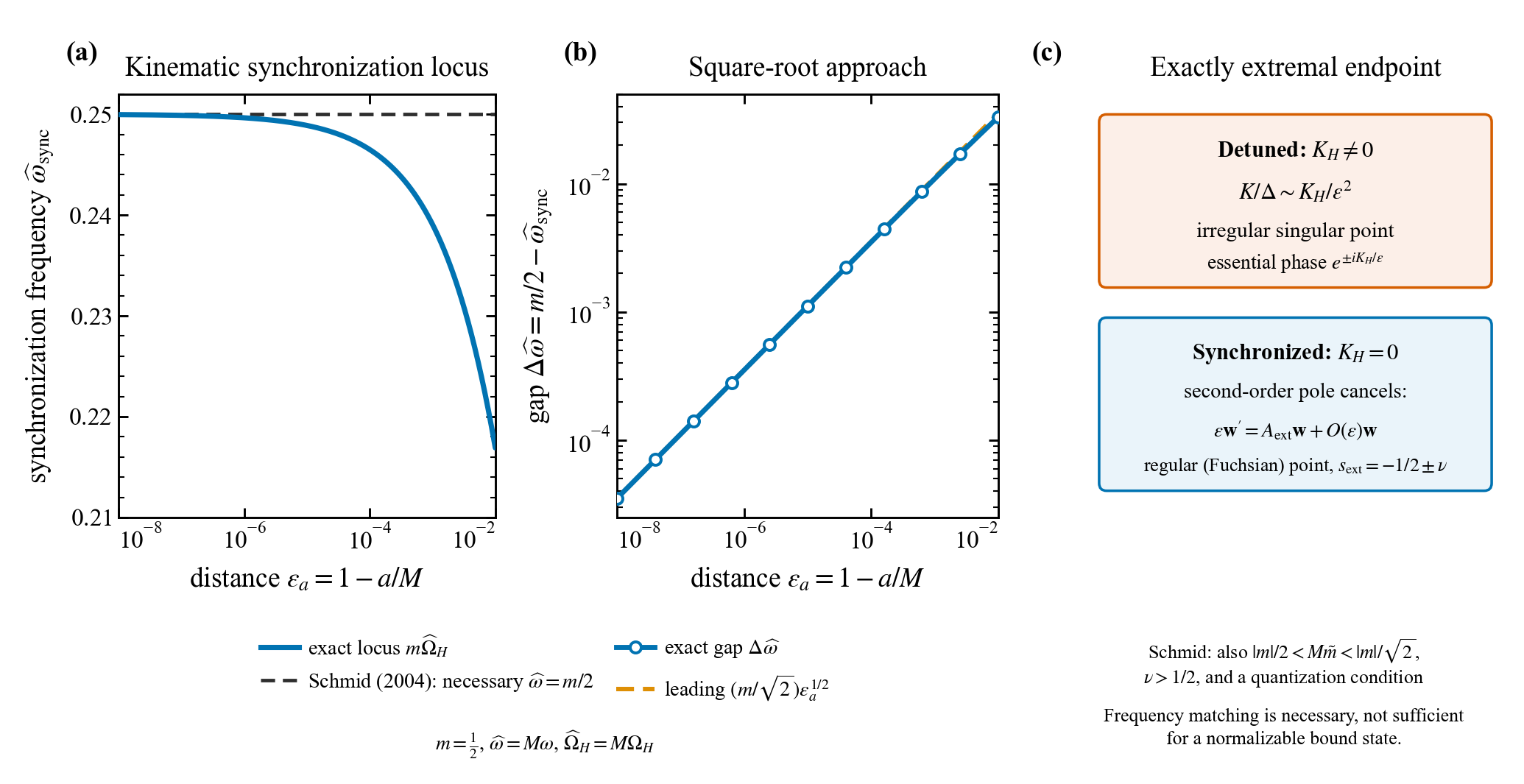}
\caption{Kinematic synchronization locus near extremality ($m=1/2$). (a) $m\Omega_{H}(a)$ approaches Schmid's frequency $m/(2M)$ (dashed). (b) The deviation $m/(2M)-m\Omega_{H}$ scales as $(m/\sqrt2)(1-a/M)^{1/2}$. (c) The exactly extremal horizon is an irregular singular point for $K_H\neq0$ and a Fuchsian point for $K_H=0$, Eq.~\eqref{eq-2.38}. This frequency is necessary but not sufficient for extremal bound states, which also require the mass window, $\nu>1/2$, and a quantization condition~\cite{Schmid:2002zf}; the curves do not represent normalizable solutions.
\label{fig:extremal_track}}
\end{figure*}

As $a\to M$ the synchronization frequency $m\Omega_H(a)$ approaches $m/(2M)$, the frequency of Schmid's extremal Dirac bound states~\cite{Schmid:2002zf}, with a deviation $\propto\sqrt{1-a/M}$ (Fig.~\ref{fig:extremal_track}). This is a kinematic locus, not a family of normalizable extremal solutions: Schmid's further conditions (Sec.~\ref{sec:2.2}) are not met, since for $|m|=1/2$ the mass window is $1/4<M\tilde m<1/(2\sqrt2)\approx0.354$, whereas $M\tilde m=0.5$ here, and the branches of Table~\ref{tab:scan} end at $a_{\rm fold}\le0.663M$. The existence of these extremal bound states has, moreover, been questioned: Batic and Nowakowski argued that Schmid's conditions admit no solution~\cite{Batic:2008hm}. None of our results depends on this question. Exactly at $a=M$ the horizon is an irregular singular point for $\omega\neq m/(2M)$ and a Fuchsian one for $\omega=m/(2M)$ [Eq.~\eqref{eq-2.38}], so the nonextremal Frobenius structure (indices $-1/4\pm ik$ in $\sqrt\varepsilon$) does not extend uniformly to $a=M$.

\par
\begin{figure*}[htbp]
\centering
\includegraphics[width=11cm]{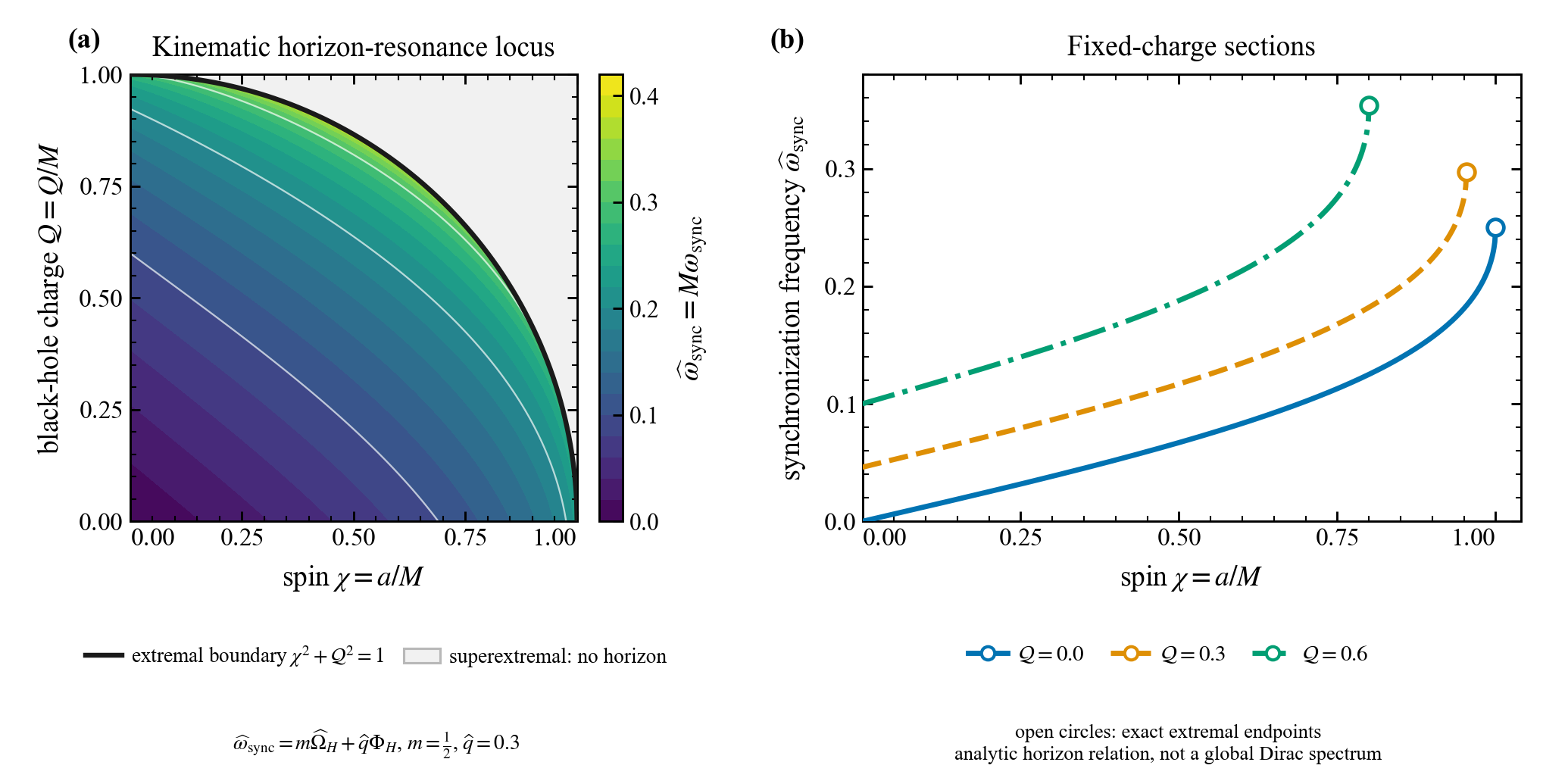}
\caption{The generalized Kerr--Newman synchronization locus for $q=0.3$. (a) Frequency $\omega_{\rm sync}=m\Omega_{H}+q\Phi_{H}$ [Eq.~\eqref{eq-4.15}] over the $(a/M,Q/M)$ plane. The black line marks the extremal boundary $a^{2}+Q^{2}=M^{2}$. (b) Slices at $Q/M=0,\,0.3,\,0.6$ show how the charge raises the synchronization frequency.
\label{fig:kn}}
\end{figure*}

Extending to the Kerr--Newman geometry via the gauge potential $A_t = -Qr/\Sigma$, the radial operator incorporates $K=(r^{2}+a^{2})\omega-am-qQr$ and $\Delta_{\rm KN}=r^{2}-2Mr+a^{2}+Q^{2}$. The horizon winding invariant incorporates the electrostatic potential $\Phi_{H}=Qr_{+}/(r_{+}^{2}+a^{2})$:
\begin{equation}
k_{\rm KN}=\frac{\omega-m\Omega_{H}-q\Phi_{H}}{2\kappa_{+}} .
\label{eq-4.14}
\end{equation}
The generalized synchronization ($k_{\rm KN}=0$) condition becomes:
\begin{equation}
\omega=m\Omega_{H}+q\Phi_{H} ,
\label{eq-4.15}
\end{equation}
matching the exact threshold for charged superradiance. (The extremal Dirac eigenfrequency $\omega=(ma-qQM)/(a^{2}+M^{2})$ of Ref.~\cite{Schmid:2002zf} differs from Eq.~\eqref{eq-4.15} only by the sign convention $q\to-q$.) Black-hole charge raises this kinematic synchronization locus (Fig.~\ref{fig:kn}); since we have not computed sourced Kerr--Newman solutions, this locus is a kinematic statement only.

\subsection{Sensitivity to the source profile}
\label{sec:4.3}
\par
\begin{figure*}[htbp]
\centering
\includegraphics[width=14cm]{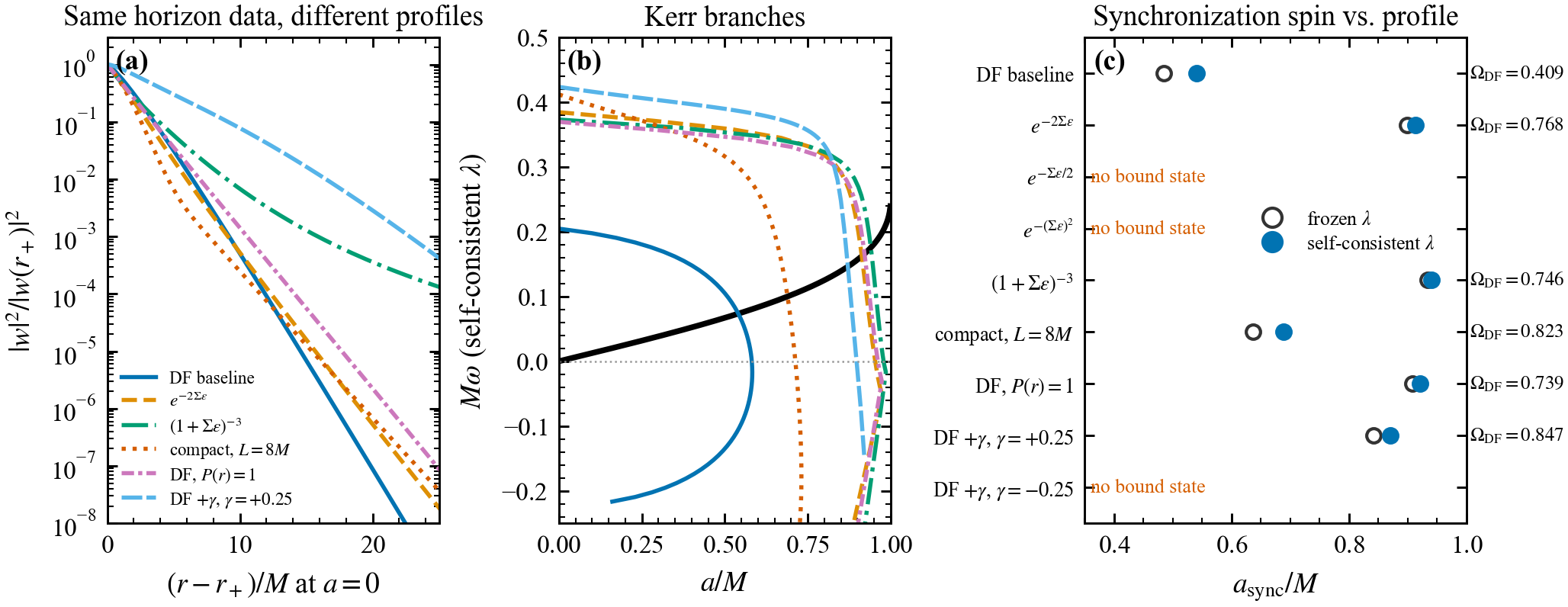}
\caption{Dependence on the postulated source profile (Case A, $v_0/u_0=-0.2$, $m=1/2$, $M\tilde m=0.5$). (a) Radial profiles $|w|^2/|w^{(0)}|^2$ at $a=0$ for the profiles with a resolved bound state; all use the same prescribed static-frame data and local matching relation. The label $w(r_+)$ in the panel denotes the exterior static-frame limit $w^{(0)}$, not a horizon value of the physical spinor. (b) Self-consistent Kerr branches $\omega(a)$ for the same profiles (line styles as in (a)) and the synchronization line $m\Omega_H$ (black). (c) $a_{\rm sync}$ for frozen (open circles) and self-consistent (filled circles) $\lambda$ for each profile, with the corresponding $\Omega_{\rm DF}(a=0)$; three profiles yield no stable positive-frequency bound-state root in the scan. The compact and $\gamma=+0.25$ branches in (b) are shown up to the spins ($a\approx0.73$ and $0.92$) at which continuation in $a$ lost them; this does not affect $a_{\rm sync}$.}
\label{fig:sens}
\end{figure*}

To separate robust results from those tied to the postulated source, we repeated the calculation for nine source profiles. All use the same prescribed leading static-frame data and the local matching relation~\eqref{eq-4.8}, but differ in the postulated parts (Sec.~\ref{sec:2.3}). The numerical source coefficients are evaluated at each resulting eigenfrequency and need not coincide between profiles. The varied ingredients are the radial envelope [the DF profile $e^{-\Sigma\varepsilon}$, $e^{-2\Sigma\varepsilon}$, $e^{-\Sigma\varepsilon/2}$, $e^{-(\Sigma\varepsilon)^2}$, $(1+\Sigma\varepsilon)^{-3}$, and the compact profile $(1-\varepsilon/L)^4$ for $\varepsilon<L=8M$], the prefactor $P(r)$ of Eq.~\eqref{eq-4.12} ($r/r_+$ or $1$), and an additional regular term $\gamma\,w^{(0)}\,(r/r_+)\,e^{-\Sigma\varepsilon}$ with $\gamma=\pm0.25$. The horizon Frobenius data are recomputed for each profile. The results are listed in Table~\ref{tab:sens} and shown in Fig.~\ref{fig:sens}.

\par
In the spherically symmetric limit the eigenfrequency depends strongly on the profile. For Case A with $v_0/u_0=-0.2$, $\Omega_{\rm DF}(a=0)$ ranges from $0.409$ (DF profile) to $0.847$, and for three profiles (the slow envelope $e^{-\Sigma\varepsilon/2}$, the Gaussian, and $\gamma=-0.25$), no stable positive-frequency bound-state root was found for these prescribed data; roots close to the threshold $|\omega|\to\tilde m$ that were not stable under changes of the matching radius were discarded. Case B behaves similarly. In Kerr, $a_{\rm sync}$ (Case A) varies between $0.48$ and $0.93$ for frozen $\lambda$ and between $0.54$ and $0.94$ for self-consistent $\lambda$, and turning points below $a=0.999$ were resolved for only some profiles. The frequencies, synchronization spins, and turning points quoted for the baseline DF profile are therefore conditional on that source prescription. They are not universal predictions of the Kerr geometry, and numerical convergence for a fixed profile does not remove this source dependence.

\par
At fixed $(a,\omega)$ and prescribed near-horizon data, the local matching relation~\eqref{eq-4.8}, its weight and phase-winding decomposition~\eqref{eq-4.9}, the index $k$, and the homogeneous Frobenius structure do not depend on the radial source envelope. The regularity and flux statements of Sec.~\ref{sec:3.2} hold for the prescribed Cases A and B, and the angular eigenproblem is independent of that envelope at fixed $(a\omega,a\tilde m)$. Its eigenvalue evaluated along a sourced branch can nevertheless change when the profile changes $\omega$. Among the tested branches that reach synchronization, the crossings are transversal and the self-consistent coupling moves $a_{\rm sync}$ to larger spins than the frozen approximation; these observations are not a theorem for arbitrary source profiles. The floor $\tfrac14+p$ of the normalized singular coefficient at $k=0$ follows directly from the local matching relation.
\begin{table*}[t]
\centering
\caption{Sensitivity to the postulated source profile ($M=1$, $m=\tfrac12$, $\tilde m=0.5$). All profiles use the same prescribed static-frame data ($v_0/u_0=-0.2$ for Case A, $-0.3$ for Case B) and the matching relation~\eqref{eq-4.8}; the numerical coefficients depend on the resulting frequency. $\Omega^+$: positive eigenfrequency at $a=0$ ($r_H$ units); $a_{\rm sync}$: synchronization spin for frozen/self-consistent $\lambda$; $P$: prefactor in Eq.~\eqref{eq-4.12}; ``none'': no stable positive-frequency bound-state root found in the scan; ``--'': not computed.
\label{tab:sens}}
\begin{tabular}{lcccc}
\hline\hline
 & \multicolumn{2}{c}{Case A} & \multicolumn{2}{c}{Case B} \\
profile & $\Omega^{+}$ & $a_{\rm sync}$ & $\Omega^{+}$ & $a_{\rm sync}$ \\
\hline
DF, $e^{-\Sigma\varepsilon}$, $P=r/r_+$ & $0.4091$ & $0.484/0.541$ & $0.7197$ & $0.768/0.817$ \\
$e^{-2\Sigma\varepsilon}$ & $0.7684$ & $0.899/0.913$ & $0.8862$ & $0.944/0.957$ \\
$e^{-\Sigma\varepsilon/2}$ & none & -- & none & -- \\
$e^{-(\Sigma\varepsilon)^2}$ & none & -- & $0.8941$ & $0.811/0.862$ \\
$(1+\Sigma\varepsilon)^{-3}$ & $0.7457$ & $0.934/0.940$ & $0.8344$ & $0.950/0.959$ \\
compact, $L=8M$ & $0.8232$ & $0.637/0.689$ & $0.9294$ & $0.856/0.892$ \\
$e^{-\Sigma\varepsilon}$, $P=1$ & $0.7386$ & $0.908/0.921$ & $0.8585$ & $0.943/0.956$ \\
DF $+\gamma$, $\gamma=+0.25$ & $0.8465$ & $0.842/0.871$ & -- & -- \\
DF $+\gamma$, $\gamma=-0.25$ & none & -- & -- & -- \\
\hline\hline
\end{tabular}
\end{table*}

\section{Conclusions}
\label{sec:5}
\par
We have studied sourced stationary Dirac bound states on Kerr backgrounds, extending the framework of Ref.~\cite{Dzhunushaliev:2026rrt}. The calculation is done in the test-field approximation with an external source: the spinor is a c-number solution of a linear sourced Dirac equation on a fixed background, and self-interactions, self-gravity, backreaction, and many-body effects are not included. We have discussed possible origins of the source and the conditions for treating it as external, and we have separated its postulated ingredients from those that follow from the field equations.

\par
Local Frobenius matching fixes the singular part of the source in terms of the leading static-frame expansion coefficient. It consists of a frame-weight part and a phase-winding part proportional to $k=(\omega-m\Omega_H)/(2\kappa_+)$, and only the latter vanishes at synchronization. In a horizon-regular frame the ingoing branch is the only regular local solution at every frequency, including synchronization, where the Dirac indicial structure has no logarithms, and an exact radial flux identity then excludes source-free normalizable stationary states. The absence of synchronized Dirac hair is thus due to the positive flux of the regular solution (the scalar flux vanishes at synchronization), not to the absence of logarithms.

\par
The sourced configurations are standing waves with zero net horizon flux, so the source does not supply a net absorption; it replaces the near-horizon behavior of the homogeneous standing waves by finite (Case A) or vanishing (Case B) static-frame amplitudes. In a horizon-regular frame Case A has an integrably divergent infalling density and no finite spinor or current value on the horizon; only the direction of its leading divergent current approaches the null generator. Case B extends continuously to zero but is not differentiable there, even at synchronization. Thus neither family is smooth on the future horizon, and the zero net horizon flux is understood as the exterior limit of the radial flux. For the baseline DF source profile, the branches connected to the Schwarzschild benchmark cross the synchronization line transversally and end, close to $\omega=0$, at a turning point where they merge with a negative-frequency branch; the self-consistent angular coupling moves both points to larger spins. The eigenfrequencies, synchronization spins, and existence and termination of the computed branches depend strongly on the postulated source profile. Profile-independent statements concern the local matching relation at fixed parameters, the kinematic synchronization locus, and the flux and regularity analysis under the stated assumptions. Transversal crossings are observed for the tested profiles whose branches reach synchronization, without implying universality for arbitrary sources.

\par
The reconstructed Kerr density is mildly oblate and does not vanish on the axis for $|m|=1/2$, whereas modes with $|m|\ge3/2$ vanish there. Toward extremality the synchronization frequency approaches Schmid's frequency $m/(2M)$, which is necessary but not sufficient for extremal bound states, and for Kerr--Newman black holes the charge shifts the synchronization locus to $\omega=m\Omega_H+q\Phi_H$. Our results rely on classical properties of the Dirac equation and do not test spin statistics. A derivation of the source from a microscopic theory and the inclusion of backreaction remain open problems.

\begin{acknowledgments}
This work is supported by the National Natural Science Foundation of China (Grant No. 12575069, 12505060, 12175027, 11875010), the Fund Project of Chongqing Normal University (Grant Number: 24XLB033), Chongqing Natural Science Foundation General Program (Grant No. CSTB2025NSCQ-GPX1019), and Science and Technology Research Project of Chongqing Municipal Education Commission (Grant Number: KJQN202500563).
\end{acknowledgments}

\vspace{1em}
\noindent \textbf{Conflict of Interest:}  The authors declare that they have no conflict of interest.

\appendix
\section{Reproducibility data}
\label{app:repro}
\par
We use $M=1$, $m=1/2$, and $\tilde m=0.5$ ($\tilde m_{\rm DF}=1$ in $r_H$ units), and prescribed static-frame coefficients $w^{(0)}=v_0+iu_0$ in Case A and $w^{(1)}=v_0+iu_0$ in Case B, with $u_0=1$; the normalization is irrelevant because the problem is linear in the data and the source.

\par
All results were obtained by integrating Eq.~\eqref{eq-4.10}, i.e., Eqs.~\eqref{eq-2.27}--\eqref{eq-2.28} with $w_2=\bar w_1$ and the operator signs of Eq.~\eqref{eq-2.15}, from $r_++\varepsilon_0$ to $r_++R_{\rm out}$ with an eighth-order Dormand--Prince integrator (relative and absolute tolerances $10^{-11}$ and $10^{-14}$), using $\lambda=\lambda_-(a\omega,a\tilde m)$ (self-consistent) or $\lambda=-1$ (frozen) with the same $\tilde m$ as in the radial equation.

\par
Case A: $\rho=(\tfrac14+ik)\,w^{(0)}$ and $\eta=-\beta_0\,\overline{w^{(0)}}$, with $\beta_0=(\lambda-i\tilde mr_+)(r_+-r_-)^{-1/2}$; Case B: $\rho'=(\tfrac34+ik)\,w^{(1)}$ and $\eta'=-\beta_0\,\overline{w^{(1)}}$. For the benchmark ($a=0$, $v_0/u_0=-0.2$, $M\omega=0.204564476$, $k=0.409128952$), $\rho=-0.459129+0.168174\,i$ and $\eta=0.565685-0.848528\,i$, i.e., $q_0+ip_0$ and $(q_1+ip_1)/\sqrt2$ of Eq.~\eqref{eq-3.4}, where $\sqrt2$ converts from $r_H=1$ to $M=1$.

\par
With $\tau=\sqrt\varepsilon$ and $w=\sum_n b_n\tau^n$, Eq.~\eqref{eq-4.10} gives the recursion
\begin{equation}
\Big(\frac n2+\alpha_0\Big)b_n=-\sum_{j\ge1}\alpha_j\,b_{n-2j}+\sum_{j\ge0}\beta_j\,\bar b_{n-1-2j}+s_n ,
\label{eq-A1}
\end{equation}
where $\varepsilon[\Delta'/(4\Delta)+iK/\Delta]=\sum_j\alpha_j\varepsilon^j$ (so that $\alpha_0=\tfrac14+ik$), $(\lambda-i\tilde mr)(r_+-r_-+\varepsilon)^{-1/2}=\sum_j\beta_j\varepsilon^j$, and $s_n$ are the coefficients of $\varepsilon J$ in powers of $\tau$. Case A imposes $b_0=w^{(0)}$ and $b_1=0$; Case B imposes $b_0=0$, $b_1=w^{(1)}$, and $b_2=0$; all other $b_n$ follow from Eq.~\eqref{eq-A1}. We use $n\le8$ and $\varepsilon_0=10^{-6}$.

\par
$S_{\pm1/2}$ are expanded in the orthonormal bases $N_jd^{\,j}_{m,\mp1/2}(\theta)$, with $N_j=\sqrt{(2j+1)/2}$ and $j=|m|,\dots,|m|+N-1$, which turns Eqs.~\eqref{eq-2.18}--\eqref{eq-2.19} into a $2N\times2N$ matrix eigenvalue problem; the eigenvectors are normalized to $\int_0^\pi(S_{+1/2}^2+S_{-1/2}^2)\sin\theta\,d\theta=1$. We use $N=40$ ($N=60$ for Figs.~\ref{fig:shape} and~\ref{fig:3d}); spectral and shooting eigenvalues agree to $\mathcal O(10^{-13})$.

\par
At large $r$ the decaying solution of Eq.~\eqref{eq-4.10} is $w\propto u_d\,e^{-\Sigma r}$ up to power-law factors, with $u_d^2=i\tilde m/(\Sigma-i\omega)$ and $|u_d|=1$. We use $\mathcal R(\omega)={\rm Im}\big[\bar u_d\,w(r_++R_{\rm out})\big]/|w(r_++R_{\rm out})|$ with $R_{\rm out}=45M$, whose zero imposes the leading asymptotic decaying direction at a finite radius, approximating the boundary condition at infinity; for $a=0$ it is equivalent to $G_X=0$ in Eq.~\eqref{eq-3.8}. Roots are bracketed on grids of $200$--$400$ points over $(-\tilde m,\tilde m)$ and refined with a Brent stopping tolerance of $10^{-14}$. This tolerance controls the root search for the computed residual, not the total eigenfrequency error from integration, truncation, and the finite outer boundary; roots near the threshold ($\Sigma R_{\rm out}\lesssim7$) are kept only if they are stable under $R_{\rm out}\to2R_{\rm out}$ and $3R_{\rm out}$.

\par
Table~\ref{tab:conv} lists the largest observed variation of $M\omega$ under the stated changes of numerical parameters for four representative configurations with the baseline DF source profile. These are empirical sensitivity checks, not rigorous error bounds or a certification of every printed digit. The extra digits in the baseline column identify the numerical output for reproducibility.

\par
For Case A at $a=0$, the largest listed change is $5\times10^{-12}$; for the two self-consistent Kerr examples it is $5\times10^{-11}$ and $4\times10^{-11}$. Case B has a larger sensitivity of $2\times10^{-8}$ when the outer boundary is reduced to $R_{\rm out}=30M$. The other listed Case B variations, including the larger outer boundaries, are below $10^{-11}$, supporting stability for those settings without certifying the last printed digits. In the spherical frequency $\Omega=2M\omega$, the corresponding largest Case B change is $4\times10^{-8}$.
\begin{table*}[t]
\centering
\caption{Observed numerical sensitivity of $M\omega$ for the baseline DF source ($v_0/u_0=-0.2$ for Case A, $-0.3$ for Case B); extra baseline digits are retained for reproducibility, not as certified accuracy. Baseline: $\varepsilon_0=10^{-6}$, $R_{\rm out}=45M$, $n\le8$, tolerance $10^{-11}$, $N=40$. Listed is the largest change under $\varepsilon_0\in\{10^{-5},10^{-7}\}$, $R_{\rm out}\in\{30,60,90\}M$, $n\le4,6,10$, tolerance $\in\{10^{-9},10^{-12}\}$, and (self-consistent cases) $N\in\{10,20,60\}$. For Case B the largest change occurs for $R_{\rm out}=30M$; all other Case B changes are below $10^{-11}$.
\label{tab:conv}}
\begin{tabular}{lcc}
\hline\hline
configuration & $M\omega$ (baseline) & max.\ change \\
\hline
$a=0$, Case A & $0.204564476190$ & $5\times10^{-12}$ \\
$a=0$, Case B & $0.359867567222$ & $2\times10^{-8}$ \\
$a=0.30$, self-cons. & $0.171904810784$ & $5\times10^{-11}$ \\
$a=0.5408$, self-cons. & $0.073428630749$ & $4\times10^{-11}$ \\
\hline\hline
\end{tabular}
\end{table*}

\par
Each arc is parametrized by $\omega$: for $79$ values of $\omega$ between the two roots at $a=0$, $\mathcal R(a;\omega)=0$ is solved for $a$ by bracketing on $a\in[0,0.95]$ (up to $0.999$ in Sec.~\ref{sec:4.3}) and Brent's method. This follows the arc through the turning point, and $a_{\rm fold}$ is obtained from a quadratic fit of $a(\omega)$ around its maximum; $a_{\rm sync}$ is found by solving $\mathcal R(a;m\Omega_H(a))=0$. Natural continuation in $a$ over the full window $-\tilde m<\omega<\tilde m$ reproduces both branches and finds no root beyond $a_{\rm fold}$. The self-consistent $\lambda$ is recomputed at every trial $(a,\omega)$. The fixed-spin frequency tests in Table~\ref{tab:conv} do not provide separate error estimates for $a_{\rm sync}$ or the fitted $a_{\rm fold}$; the latter also depends on the continuation sampling and fit. Nor do those tests establish uniform accuracy near the mass threshold or for all source profiles in Sec.~\ref{sec:4.3}, or prove completeness of the root search.

\par
For the configurations at $a=0$, $0.30$, and $0.5408$ ($v_0/u_0=-0.2$) and at $a=0.45$ ($v_0/u_0=-0.3$), the full two-component system with $\mathcal J_2=\bar{\mathcal J}_1$ preserves $|w_1|=|w_2|$, and $\mathfrak F$ and the integrated creation rate vanish to round-off. For data off the subspace the two sides of Eq.~\eqref{eq-flux} agree to $4\times10^{-8}$ or better, and for the homogeneous ingoing branch $\mathfrak F$ is constant to seven digits.

\par
With computer algebra we have verified that the operators of Eq.~\eqref{eq-2.15} reproduce the Teukolsky equations for $s=\mp1/2$ when $\tilde m=0$ (the opposite sign of $iK/\Delta$ does not), that the map~\eqref{eq-4.11} turns Eq.~\eqref{eq-4.10} into Eqs.~\eqref{eq-3.1}--\eqref{eq-3.2}, that the residues~\eqref{eq-4.8} reproduce Eqs.~\eqref{eq-3.4}--\eqref{eq-3.5}, and the expressions~\eqref{eq-2.6} and~\eqref{eq-scalarflux}. The scripts used for all figures and tables are available from the corresponding authors upon reasonable request.



\nocite{*}

\bibliography{apssamp}

@article{Hawking:1975vcx,
    author = "Hawking, S. W.",
    title = "{Particle Creation by Black Holes}",
    doi = "10.1007/BF02345020",
    journal = "Commun. Math. Phys.",
    volume = "43",
    pages = "199--220",
    year = "1975",
    note = "[Erratum: Commun. Math. Phys. 46, 206 (1976)]"
}

@article{Bekenstein:1973ur,
    author = "Bekenstein, Jacob D.",
    title = "{Black holes and entropy}",
    doi = "10.1103/PhysRevD.7.2333",
    journal = "Phys. Rev. D",
    volume = "7",
    pages = "2333--2346",
    year = "1973"
}

@article{Bardeen:1973gs,
    author = "Bardeen, James M. and Carter, B. and Hawking, S. W.",
    title = "{The four laws of black hole mechanics}",
    doi = "10.1007/BF01645742",
    journal = "Commun. Math. Phys.",
    volume = "31",
    pages = "161--170",
    year = "1973"
}

@article{Maggiore:1993kv,
    author = "Maggiore, Michele",
    title = "{The algebraic structure of the generalized uncertainty principle}",
    eprint = "hep-th/9309034",
    archivePrefix = "arXiv",
    reportNumber = "IFUP-TH-38-93",
    doi = "10.1016/0370-2693(93)90785-G",
    journal = "Phys. Lett. B",
    volume = "319",
    pages = "83--86",
    year = "1993"
}

@article{Capozziello:1999wx,
    author = "Capozziello, S. and Lambiase, G. and Scarpetta, G.",
    title = "{Generalized uncertainty principle from quantum geometry}",
    eprint = "gr-qc/9910017",
    archivePrefix = "arXiv",
    doi = "10.1023/A:1003634814685",
    journal = "Int. J. Theor. Phys.",
    volume = "39",
    pages = "15--22",
    year = "2000"
}

@article{Chandrasekhar:1976ap,
    author = "Chandrasekhar, Subrahmanyan",
    title = "{The Solution of Dirac's Equation in Kerr Geometry}",
    doi = "10.1098/rspa.1976.0090",
    journal = "Proc. Roy. Soc. Lond. A",
    volume = "349",
    pages = "571--575",
    year = "1976"
}

@article{Page:1976jj,
    author = "Page, Don N.",
    title = "{Dirac Equation Around a Charged, Rotating Black Hole}",
    doi = "10.1103/PhysRevD.14.1509",
    journal = "Phys. Rev. D",
    volume = "14",
    pages = "1509--1510",
    year = "1976"
}

@book{Chandrasekhar:1984siy,
    author    = "Chandrasekhar, S.",
    title     = "{The Mathematical Theory of Black Holes}",
    publisher = "Clarendon Press",
    address   = "Oxford",
    series    = "International Series of Monographs on Physics",
    volume    = "69",
    year      = "1983",
    isbn      = "0-19-851291-0"
}

@article{Finster:1998ws,
    author = "Finster, Felix and Smoller, Joel and Yau, Shing-Tung",
    title = "{Particlelike solutions of the Einstein-Dirac equations}",
    eprint = "gr-qc/9801079",
    archivePrefix = "arXiv",
    doi = "10.1103/PhysRevD.59.104020",
    journal = "Phys. Rev. D",
    volume = "59",
    pages = "104020",
    year = "1999"
}

@article{Herdeiro:2017fhv,
    author = "Herdeiro, Carlos A. R. and Pombo, Alexandre M. and Radu, Eugen",
    title = "{Asymptotically flat scalar, Dirac and Proca stars: discrete vs. continuous families of solutions}",
    eprint = "1708.05674",
    archivePrefix = "arXiv",
    primaryClass = "gr-qc",
    doi = "10.1016/j.physletb.2017.09.036",
    journal = "Phys. Lett. B",
    volume = "773",
    pages = "654--662",
    year = "2017"
}

@article{Herdeiro:2019mbz,
    author = "Herdeiro, C. and Perapechka, I. and Radu, E. and Shnir, {\relax Ya}.",
    title = "{Asymptotically flat spinning scalar, Dirac and Proca stars}",
    eprint = "1906.05386",
    archivePrefix = "arXiv",
    primaryClass = "gr-qc",
    doi = "10.1016/j.physletb.2019.134845",
    journal = "Phys. Lett. B",
    volume = "797",
    pages = "134845",
    year = "2019"
}

@article{Dzhunushaliev:2025dma,
    author = "Dzhunushaliev, Vladimir and Folomeev, Vladimir",
    title = "{Condensation of a spinor field at the event horizon}",
    eprint = "2505.22163",
    archivePrefix = "arXiv",
    primaryClass = "gr-qc",
    doi = "10.1016/j.physletb.2026.140416",
    journal = "Phys. Lett. B",
    volume = "876",
    pages = "140416",
    year = "2026"
}

@article{Finster:1999ry,
    author = "Finster, Felix and Kamran, Niky and Smoller, Joel and Yau, Shing-Tung",
    title = "{Nonexistence of time-periodic solutions of the Dirac equation in an axisymmetric black hole geometry}",
    eprint = "gr-qc/9905047",
    archivePrefix = "arXiv",
    doi = "10.1002/(SICI)1097-0312(200007)53:7<902::AID-CPA4>3.0.CO;2-4",
    journal = "Commun. Pure Appl. Math.",
    volume = "53",
    pages = "902--929",
    year = "2000",
    note = "[Erratum: Commun. Pure Appl. Math. 53, 1201 (2000)]"
}

@article{Finster:2000jz,
    author = "Finster, Felix and Kamran, Niky and Smoller, Joel and Yau, Shing-Tung",
    title = "{The long-time dynamics of Dirac particles in the Kerr-Newman black hole geometry}",
    eprint = "gr-qc/0005088",
    archivePrefix = "arXiv",
    doi = "10.4310/ATMP.2003.v7.n1.a2",
    journal = "Adv. Theor. Math. Phys.",
    volume = "7",
    number = "1",
    pages = "25--52",
    year = "2003"
}

@article{Schmid:2002zf,
    author        = "Schmid, Harald",
    title         = "{Bound state solutions of the Dirac equation in the extreme Kerr geometry}",
    journal       = "Math. Nachr.",
    volume        = "274--275",
    number        = "1",
    pages         = "117--129",
    year          = "2004",
    doi           = "10.1002/mana.200410205",
    eprint        = "math-ph/0207039",
    archivePrefix = "arXiv"
}

@article{Hod:2012px,
    author        = "Hod, Shahar",
    title         = "{Stationary scalar clouds around rotating black holes}",
    journal       = "Phys. Rev. D",
    volume        = "86",
    number        = "10",
    pages         = "104026",
    year          = "2012",
    note          = "[Erratum: Phys. Rev. D 86, 129902 (2012)]",
    doi           = "10.1103/PhysRevD.86.104026",
    eprint        = "1211.3202",
    archivePrefix = "arXiv",
    primaryClass  = "gr-qc"
}

@article{Herdeiro:2014goa,
    author = "Herdeiro, Carlos A. R. and Radu, Eugen",
    title = "{Kerr black holes with scalar hair}",
    eprint = "1403.2757",
    archivePrefix = "arXiv",
    primaryClass = "gr-qc",
    doi = "10.1103/PhysRevLett.112.221101",
    journal = "Phys. Rev. Lett.",
    volume = "112",
    pages = "221101",
    year = "2014"
}

@misc{Guo:2026pdi,
    author = "Guo, Sen and Wen, Lin and Zeng, Xiao-Xiong",
    title = "{Superradiant Bose--Einstein condensates around Kerr black holes}",
    eprint = "2608.02051",
    archivePrefix = "arXiv",
    primaryClass = "gr-qc",
    month = "8",
    year = "2026"
}

@misc{Dzhunushaliev:2026rrt,
    author = "Dzhunushaliev, Vladimir and Folomeev, Vladimir",
    title = "{Fermion condensate at the event horizon}",
    eprint = "2605.21064",
    archivePrefix = "arXiv",
    primaryClass = "gr-qc",
    month = "5",
    year = "2026"
}

@article{McMaken:2024tpc,
    author = "McMaken, Tyler and Hamilton, Andrew J. S.",
    title = "{Hawking radiation inside a rotating black hole}",
    eprint = "2401.03098",
    archivePrefix = "arXiv",
    primaryClass = "gr-qc",
    doi = "10.1103/PhysRevD.109.065023",
    journal = "Phys. Rev. D",
    volume = "109",
    number = "6",
    pages = "065023",
    year = "2024"
}

@article{Dubey:2025hwk,
    author = "Dubey, Nitesh Kumar and Kolekar, Sanved",
    title = "{Harvesting fermionic field entanglement in Schwarzschild spacetime}",
    eprint = "2504.06887",
    archivePrefix = "arXiv",
    primaryClass = "gr-qc",
    reportNumber = "112",
    doi = "10.1103/1mwx-7jmf",
    journal = "Phys. Rev. D",
    volume = "112",
    number = "2",
    pages = "025019",
    year = "2025"
}

@book{Wald:1984rg,
    author = "Wald, Robert M.",
    title = "{General Relativity}",
    doi = "10.7208/chicago/9780226870373.001.0001",
    publisher = "University of Chicago Press",
    address = "Chicago",
    year = "1984"
}

@article{Fewster:2025vxz,
    author = "Fewster, Christopher J.",
    title = "{Hadamard States for Decomposable Green-Hyperbolic Operators}",
    eprint = "2503.12537",
    archivePrefix = "arXiv",
    primaryClass = "math-ph",
    doi = "10.1007/s00220-025-05512-1",
    journal = "Commun. Math. Phys.",
    volume = "407",
    number = "1",
    pages = "14",
    year = "2026"
}

@article{Kerr:1963ud,
    author = "Kerr, Roy P.",
    title = "{Gravitational field of a spinning mass as an example of algebraically special metrics}",
    doi = "10.1103/PhysRevLett.11.237",
    journal = "Phys. Rev. Lett.",
    volume = "11",
    pages = "237--238",
    year = "1963"
}

@book{Misner:1973prb,
    author = "Misner, Charles W. and Thorne, K. S. and Wheeler, J. A.",
    title = "{Gravitation}",
    isbn = "978-0-7167-0344-0, 978-0-691-17779-3",
    publisher = "W. H. Freeman",
    address = "San Francisco",
    year = "1973"
}

@article{Zeldovich:1971ffh,
    author = "Zel'dovich, {\relax Ya}. B.",
    title = "{Generation of Waves by a Rotating Body}",
    journal = "JETP Lett.",
    volume = "14",
    pages = "180",
    year = "1971",
    note = "[Pis'ma Zh. Eksp. Teor. Fiz. 14, 270 (1971)]"
}

@article{Konoplya:2021hsm,
    author = "Konoplya, R. A. and Zhidenko, A.",
    title = "{Traversable Wormholes in General Relativity}",
    eprint = "2106.05034",
    archivePrefix = "arXiv",
    primaryClass = "gr-qc",
    doi = "10.1103/PhysRevLett.128.091104",
    journal = "Phys. Rev. Lett.",
    volume = "128",
    number = "9",
    pages = "091104",
    year = "2022"
}

@article{Dzhunushaliev:2025ntr,
    author = "Dzhunushaliev, Vladimir and Folomeev, Vladimir",
    title = "{Rotating wormholes in Einstein-Dirac-Maxwell theory}",
    eprint = "2511.02431",
    archivePrefix = "arXiv",
    primaryClass = "gr-qc",
    doi = "10.1007/s10714-026-03550-1",
    journal = "Gen. Rel. Grav.",
    volume = "58",
    number = "4",
    pages = "39",
    year = "2026"
}

@article{Dzhunushaliev:2025lki,
    author = "Dzhunushaliev, Vladimir and Folomeev, Vladimir and Beissen, Nurzada and Nurmukhamedov, Adilet",
    title = "{Wormholes in Einstein-Dirac-Maxwell theory with identical spacetime asymptotics}",
    eprint = "2510.03656",
    archivePrefix = "arXiv",
    primaryClass = "gr-qc",
    doi = "10.1140/epjc/s10052-026-15456-y",
    journal = "Eur. Phys. J. C",
    volume = "86",
    number = "3",
    pages = "240",
    year = "2026"
}

@article{Brito:2015rjv,
    author = "Brito, Richard and Cardoso, Vitor and Pani, Paolo",
    title = "{Superradiance: Energy Extraction, Black-Hole Bombs and Implications for Astrophysics and Particle Physics}",
    doi = "10.1007/978-3-319-19000-6",
    journal = "Lect. Notes Phys.",
    volume = "906",
    pages = "1--237",
    year = "2015",
    eprint = "1501.06570",
    archivePrefix = "arXiv",
    primaryClass = "gr-qc"
}

@article{Finster:1998ak,
    author        = "Finster, Felix and Smoller, Joel and Yau, Shing-Tung",
    title         = "{Non-existence of time-periodic solutions of the Dirac equation in a Reissner-Nordstr{\"o}m black hole background}",
    journal       = "J. Math. Phys.",
    volume        = "41",
    number        = "4",
    pages         = "2173--2194",
    year          = "2000",
    doi           = "10.1063/1.533234",
    eprint        = "gr-qc/9805050",
    archivePrefix = "arXiv"
}

@article{Finster:1998ju,
    author        = "Finster, Felix and Smoller, Joel and Yau, Shing-Tung",
    title         = "{Non-existence of black hole solutions for a spherically symmetric, static Einstein-Dirac-Maxwell system}",
    journal       = "Commun. Math. Phys.",
    volume        = "205",
    number        = "2",
    pages         = "249--262",
    year          = "1999",
    doi           = "10.1007/s002200050675",
    eprint        = "gr-qc/9810048",
    archivePrefix = "arXiv"
}

@article{Lasenby:2002mc,
    author        = "Lasenby, Anthony and Doran, Chris and Pritchard, Jonathan and Caceres, Alejandro and Dolan, Sam",
    title         = "{Bound states and decay times of fermions in a Schwarzschild black hole background}",
    journal       = "Phys. Rev. D",
    volume        = "72",
    number        = "10",
    pages         = "105014",
    year          = "2005",
    doi           = "10.1103/PhysRevD.72.105014",
    eprint        = "gr-qc/0209090",
    archivePrefix = "arXiv"
}

@article{Dolan:2015eua,
    author        = "Dolan, Sam R. and Dempsey, David",
    title         = "{Bound states of the Dirac equation on Kerr spacetime}",
    journal       = "Class. Quant. Grav.",
    volume        = "32",
    number        = "18",
    pages         = "184001",
    year          = "2015",
    doi           = "10.1088/0264-9381/32/18/184001",
    eprint        = "1504.03190",
    archivePrefix = "arXiv",
    primaryClass  = "gr-qc"
}

@book{Birrell:1982ix,
    author    = "Birrell, N. D. and Davies, P. C. W.",
    title     = "{Quantum Fields in Curved Space}",
    publisher = "Cambridge University Press",
    address   = "Cambridge, UK",
    series    = "Cambridge Monographs on Mathematical Physics",
    year      = "1982",
    doi       = "10.1017/CBO9780511622632"
}

@article{Teukolsky:1974yv,
    author  = "Teukolsky, Saul A. and Press, William H.",
    title   = "{Perturbations of a rotating black hole. III. Interaction of the hole with gravitational and electromagnetic radiation}",
    journal = "Astrophys. J.",
    volume  = "193",
    pages   = "443--461",
    year    = "1974",
    doi     = "10.1086/153180"
}

@article{Unruh:1974bw,
    author  = "Unruh, W. G.",
    title   = "{Second quantization in the Kerr metric}",
    journal = "Phys. Rev. D",
    volume  = "10",
    number  = "10",
    pages   = "3194--3205",
    year    = "1974",
    doi     = "10.1103/PhysRevD.10.3194"
}

@article{Martellini:1977qf,
    author  = "Martellini, M. and Treves, A.",
    title   = "{Absence of superradiance of a Dirac field in a Kerr background}",
    journal = "Phys. Rev. D",
    volume  = "15",
    number  = "10",
    pages   = "3060--3061",
    year    = "1977",
    doi     = "10.1103/PhysRevD.15.3060"
}

@book{CoddingtonLevinson:1955,
    author    = "Coddington, Earl A. and Levinson, Norman",
    title     = "{Theory of Ordinary Differential Equations}",
    publisher = "McGraw-Hill",
    address   = "New York",
    series    = "International Series in Pure and Applied Mathematics",
    year      = "1955"
}

@book{Wasow:1965,
    author    = "Wasow, Wolfgang",
    title     = "{Asymptotic Expansions for Ordinary Differential Equations}",
    publisher = "Interscience Publishers",
    address   = "New York",
    series    = "Pure and Applied Mathematics",
    volume    = "14",
    year      = "1965"
}

@article{Teukolsky:1973ha,
    author  = "Teukolsky, Saul A.",
    title   = "{Perturbations of a rotating black hole. I. Fundamental equations for gravitational, electromagnetic, and neutrino-field perturbations}",
    journal = "Astrophys. J.",
    volume  = "185",
    pages   = "635--647",
    year    = "1973",
    doi     = "10.1086/152444"
}

@article{Berti:2005gp,
    author        = "Berti, Emanuele and Cardoso, Vitor and Casals, Marc",
    title         = "{Eigenvalues and eigenfunctions of spin-weighted spheroidal harmonics in four and higher dimensions}",
    journal       = "Phys. Rev. D",
    volume        = "73",
    number        = "2",
    pages         = "024013",
    year          = "2006",
    note          = "[Erratum: Phys. Rev. D 73, 109902 (2006)]",
    doi           = "10.1103/PhysRevD.73.024013",
    eprint        = "gr-qc/0511111",
    archivePrefix = "arXiv"
}

@article{Hod:2008zz,
    author        = "Hod, Shahar",
    title         = "{Slow relaxation of rapidly rotating black holes}",
    journal       = "Phys. Rev. D",
    volume        = "78",
    number        = "8",
    pages         = "084035",
    year          = "2008",
    doi           = "10.1103/PhysRevD.78.084035",
    eprint        = "0811.3806",
    archivePrefix = "arXiv",
    primaryClass  = "gr-qc"
}

@article{Yang:2013uba,
    author        = "Yang, Huan and Zimmerman, Aaron and Zengino{\u{g}}lu, An{\i}l and Zhang, Fan and Berti, Emanuele and Chen, Yanbei",
    title         = "{Quasinormal modes of nearly extremal Kerr spacetimes: spectrum bifurcation and power-law ringdown}",
    journal       = "Phys. Rev. D",
    volume        = "88",
    number        = "4",
    pages         = "044047",
    year          = "2013",
    doi           = "10.1103/PhysRevD.88.044047",
    eprint        = "1307.8086",
    archivePrefix = "arXiv",
    primaryClass  = "gr-qc"
}

@book{Peskin:1995ev,
    author    = "Peskin, Michael E. and Schroeder, Daniel V.",
    title     = "{An Introduction to Quantum Field Theory}",
    publisher = "Addison-Wesley",
    address   = "Reading, MA",
    year      = "1995",
    isbn      = "978-0-201-50397-5"
}

@article{Hawking:1972hy,
    author  = "Hawking, S. W. and Hartle, J. B.",
    title   = "{Energy and angular momentum flow into a black hole}",
    journal = "Commun. Math. Phys.",
    volume  = "27",
    number  = "4",
    pages   = "283--290",
    year    = "1972",
    doi     = "10.1007/BF01645515"
}

@article{Geroch:1973am,
    author  = "Geroch, R. and Held, A. and Penrose, R.",
    title   = "{A space-time calculus based on pairs of null directions}",
    journal = "J. Math. Phys.",
    volume  = "14",
    number  = "7",
    pages   = "874--881",
    year    = "1973",
    doi     = "10.1063/1.1666410"
}

@article{Bardeen:1972fi,
    author  = "Bardeen, James M. and Press, William H. and Teukolsky, Saul A.",
    title   = "{Rotating black holes: locally nonrotating frames, energy extraction, and scalar synchrotron radiation}",
    journal = "Astrophys. J.",
    volume  = "178",
    pages   = "347--369",
    year    = "1972",
    doi     = "10.1086/151796"
}

@article{Dolan:2009kj,
    author        = "Dolan, Sam R. and Gair, Jonathan R.",
    title         = "{The massive Dirac field on a rotating black hole spacetime: angular solutions}",
    journal       = "Class. Quant. Grav.",
    volume        = "26",
    number        = "17",
    pages         = "175020",
    year          = "2009",
    doi           = "10.1088/0264-9381/26/17/175020",
    eprint        = "0905.2974",
    archivePrefix = "arXiv",
    primaryClass  = "gr-qc"
}

@article{Jing:2005pk,
    author        = "Jing, Jiliang and Pan, Qiyuan",
    title         = "{Dirac quasinormal frequencies of the Kerr-Newman black hole}",
    journal       = "Nucl. Phys. B",
    volume        = "728",
    pages         = "109--120",
    year          = "2005",
    doi           = "10.1016/j.nuclphysb.2005.08.038",
    eprint        = "gr-qc/0506098",
    archivePrefix = "arXiv"
}

@article{Carter:1968rr,
    author  = "Carter, Brandon",
    title   = "{Global structure of the Kerr family of gravitational fields}",
    journal = "Phys. Rev.",
    volume  = "174",
    number  = "5",
    pages   = "1559--1571",
    year    = "1968",
    doi     = "10.1103/PhysRev.174.1559"
}

@article{Batic:2008hm,
    author        = "Batic, Davide and Nowakowski, Marek",
    title         = "{On the bound states of the Dirac equation in the extreme Kerr metric}",
    journal       = "Class. Quant. Grav.",
    volume        = "25",
    number        = "22",
    pages         = "225022",
    year          = "2008",
    doi           = "10.1088/0264-9381/25/22/225022",
    eprint        = "0805.4828",
    archivePrefix = "arXiv",
    primaryClass  = "gr-qc"
}

@CONTROL{REVTEX42Control}

@CONTROL{apsrev42Control,author="08",editor="1",pages="0",title="0",year="1"}



\end{document}